\documentclass[sn-nature]{sn-jnl} 

\usepackage{booktabs} 
\usepackage{siunitx}
\usepackage{amsmath}
\usepackage{times}  
\usepackage{setspace}
\usepackage{graphicx}
\usepackage{subcaption} 

\usepackage{float} 

\usepackage{tabularx}
\usepackage{threeparttable}

\begin{document}

\title{Process–Microstructure Coupling in Reduced Gravity Laser Welding via Open-Source Multiphysics Simulation Framework}

\author[1]{\fnm{Rakibul Islam} \sur{Kanak}}
\author[1]{\fnm{Taslima Hossain} \sur{Sanjana}}
\author[2]{\fnm{Apurba} \sur{Sarker}}
\author*[2]{\fnm{Sourav} \sur{Saha}}\email{souravsaha@vt.edu}

\affil[1]{\orgdiv{Department of Mechanical Engineering}, \orgname{Bangladesh University of Engineering and Technology}, \orgaddress{\city{Dhaka}, \country{Bangladesh}}}
\affil[2]{\orgdiv{Kevin T. Crofton Department of Aerospace and Ocean Engineering}, \orgname{Virginia Polytechnic Institute and State University}, \orgaddress{\city{1600 Innovation Drive, Blacksburg, Virginia}, \country{USA, 24061}}}



\abstract{
Supplying spare parts from Earth for in-space repair is economically prohibitive and logistically slow, posing a major barrier to sustainable space operations. As lunar and Martian missions accelerate in the coming decades, the feasibility of in-situ repair methods—particularly laser-based welding—must be rigorously evaluated. The micro-scale physics governing weld quality are fundamentally altered by variations in gravity and ambient pressure, yet their coupled influence across different welding regimes remains poorly understood. This work introduces a fully open-source thermo-fluid-microstructure modeling framework with computational fluid dynamics (CFD) and cellular automata (CA) to quantify how gravitational conditions reshape weld pool behavior across multiple welding regimes and spatial scales. The framework further enables prediction of the resulting microstructure from imposed process parameters, providing an accessible pathway for future research in in-space manufacturing (ISM). As a demonstration, the article analyzes laser welding of Al6061 across three process regimes (conduction, transition, and keyhole) and four gravitational environments (Earth, Moon, Mars, and International Space Station (ISS)). Analysis reveals that reduced gravity suppresses buoyancy-driven convection, thereby altering melt pool geometries. Vacuum conditions increase laser energy deposition, while microgravity promotes equiaxed grain formation via reduced melt convection. The framework captures keyhole dynamics, thermal histories, and grain morphologies with high fidelity, validated against experimental data. These findings establish process–microstructure relationships critical for reliable metallic fabrication in space and provide a reproducible platform for future in-space manufacturing research.
}


\maketitle
\section{Introduction}
At roughly \$2,720 per kilogram to low Earth orbit, even small payloads carry a staggering price tag—turning routine maintenance of in-space components into a logistical and financial burden \cite{jonesRecentLargeReduction}. This logistical and financial barrier necessitates successful repair or manufacturing of critical components in space, i.e., efficient in-space manufacturing (ISM). In-space welding (ISW), more precisely, Laser Beam Welding (LBW) has proven to be a promising candidate for ISM. Its vacuum operability and ease of robotic integration make it particularly appropriate for extraterrestrial applications \cite{klimpelReviewAnalysisModern2024,choReviewLaserWelding2024}. Furthermore, in comparison with conventional arc welding, the deep and narrow fusion zone in LBW produces a much smaller heat-affected zone, resulting in less thermal distortion and superior post-weld mechanical properties \cite{kalaiselvanRoutesJoiningMetal2021}.

The thermofluid mechanics that govern the molten pool, keyhole dynamics, and solidification process during laser welding alter significantly in space due to the combined effects of microgravity, extreme temperature variations, enhanced radiation, and high-vacuum conditions \cite{nunesLOWPRESSUREGAS2002}. The suppression of buoyancy-driven convection alters weld pool geometry and increases bubble entrapment \cite{luoInfluenceGravityState2013, aoklWeldFormationMicrogravity, 3DPrintingSpace2014}. Space environments outside the station constitute a hard vacuum (potentially $10^{-19}$ Pa), while locations like the lunar surface maintain a high vacuum ($10^{-9}$ Pa) \cite{nunesLOWPRESSUREGAS2002}. These high-vacuum environments suppress the laser-induced metal vapor plume \cite{choReviewLaserWelding2024}, leading to reduced laser attenuation and scattering, thereby delivering more laser power to the work surface \cite{brownAtmosphereEffectsLaser2024}. Reduced gravity not only impacts the keyhole and vapor dynamics but also fundamentally changes microstructural behavior. In a benchmark microgravity directional solidification study, researchers found that reduced melt convection significantly influences microstructural evolution. $Al-Cu$ alloys solidified under microgravity formed fully equiaxed grains, whereas comparable Earth-grown samples developed columnar grains, demonstrating the columnar-to-equiaxed transition (CET) \cite{williamsBenchmarkAlCuSolidification2023}. However, microstructural evolution during laser welding may differ substantially due to variations in cooling rate and process conditions.

Research has long established that laser welding of aluminum alloys (such as AA6061, AA5083, and AA2219) is inherently challenging due to their high reflectivity, high thermal conductivity, low viscosity, and susceptibility to defects like porosity and thermal cracking \cite{fayeEffectsWeldingParameters2021, dakkaInfluenceBeamOscillation2025}. Numerous terrestrial studies have pursued optimization strategies for Al6061 laser welding, including the use of grain refiners \cite{carluccioGrainRefinementLaser2018} for microstructure control, nanoparticles to enhance energy absorption \cite{quNanoparticleenabledIncreaseEnergy2022}, and ultrasonic treatment to modify melt pool behavior \cite{mutswatiwaHighspeedSynchrotronXray2024}. Some experimental investigations have manipulated the welding atmosphere itself to address these challenges. A study on vacuum laser beam welding (VLBW) of aluminum alloys demonstrated that reducing ambient pressure from atmospheric levels (1000 mbar) to 10 mbar significantly improves weld quality by suppressing the plasma plume and enabling more efficient laser energy transfer to the substrate. Experiments on Al6061-T6 alloy showed that at 10 mbar, penetration depth increased by up to 147.8\%, while porosity decreased from 12.1\% to 0.28\% \cite{leeVacuumLaserBeam2021}. However, the situation becomes more complex in space, where the absence of gravity can increase gas entrapment due to suppressed buoyancy forces, potentially exacerbating porosity formation. Therefore, the combined influence of vacuum conditions and microgravity on the laser weldability and microstructural behavior of Al6061---a material used abundantly for its high strength-to-weight ratio---remains unexplored. The only known microgravity study on aluminum microstructure, the 1977 Skylab M551 Metals Melting experiment on 2219-T87 Al, found that zero-gravity electron beam welds exhibited a central equiaxed grain zone, coarser dendrites, increased Cu segregation, and hot tearing absent in terrestrial welds \cite{LiMetalsMelting1976}. 

Given the limitations and considerable expense associated with extensive physical testing across multiple extraterrestrial environments \cite{choReviewLaserWelding2024}, high-fidelity computational methods are indispensable for investigating these complex, coupled physical mechanisms. Recent advances in simulating vacuum-specific effects during laser keyhole welding \cite{hanStudyLaserKeyhole2021} and gravity effects in laser metal deposition \cite{guComputationalFluidDynamic2019} demonstrate promising capabilities. Microstructure simulation approaches, such as Monte Carlo (MC) \cite{yangThreeDimensionalMonte2000}, phase field (PF) \cite{boettingerPhaseFieldSimulationSolidification2002}, and cellular automata (CA) \cite{gandinCoupledFiniteElementcellular1994}, can link thermal histories to grain evolution, with the CA method being particularly well-suited for part-scale microstructure prediction at lower computational costs. However, most existing computational frameworks for metal additive manufacturing and welding rely on proprietary software, limiting accessibility and reproducibility for the broader research community. The development of open-source tools democratizes access to high-fidelity simulation capabilities, enabling researchers worldwide---particularly those in resource-constrained institutions---to contribute to advancing space manufacturing technologies without prohibitive licensing costs.

To our knowledge, this is the first systematic study examining how microgravity and vacuum jointly influence melt pool dynamics and microstructural evolution of Al6061 across different welding modes using a fully open-source computational framework. This study investigates the influence of varying gravitational environments and welding modes on the resulting meltpool size, morphology, and grain textures during laser welding. To achieve this, a one-way coupled, high-fidelity computational fluid dynamics-cellular automata (CFD-CA) model was employed. The keyhole dynamics and thermal history were simulated using LaserbeamFOAM, an open-source finite volume, ray-traced, multiphase Volume of Fluid (VOF) solver \cite{flintLaserbeamFoamLaserRaytracing2023}. The resulting thermal data was then used as input for the open-source CA solver ExaCA to simulate grain growth \cite{rolchigoExaCAPerformancePortable2022}. The investigation encompasses twelve distinct cases, simulating three modes of welding (conduction, transition, and keyhole) across four gravitational and pressure conditions: Earth, Mars, the Moon, and the International Space Station (ISS). The welding modes were defined based on the Keyhole Number, as established by Gan et al. \cite{ganUniversalScalingLaws2021}. By releasing the complete workflow, input parameters, and post-processing scripts as open-source resources, this work enables transparent validation, reproducibility, and extension by the global research community. Ultimately, this research provides both the fundamental understanding and accessible computational infrastructure necessary to establish process guidelines for reliably fabricating and repairing metallic structures in space, paving the way for sustainable extraterrestrial manufacturing.

\section{Methods}

\subsection{Thermo-Fluid Modeling}

The laser welding process is modeled using LaserbeamFoam, an open-source thermo-fluid solver developed within the OpenFOAM-10 C++ Finite Volume framework \cite{flintLaserbeamFoamLaserRaytracing2023}. This solver employs the multiphase Volume of Fluid (VOF) method to accurately capture the sharp interface between the metallic substrate and the surrounding gaseous domain. A ray-tracing algorithm based on Fresnel absorption equations is used to model the laser material interaction with high fidelity.




To predict the transient temperature and velocity fields in the melt pool region, three sets of conservation equations are solved, which are conservation of mass, momentum and energy equations. The mass conservation equation is expressed as,

\begin{equation}
\nabla \cdot \mathbf{U} = 0,
\label{eq:mass}
\end{equation}

\noindent where $\mathbf{U}$ is the velocity. The momentum conservation equation is expressed as follows, 

\begin{equation}
\frac{\partial(\rho \mathbf{U})}{\partial t} 
+ \nabla\cdot(\rho \mathbf{U}\otimes \mathbf{U}) 
= -\nabla P + \nabla\cdot\boldsymbol{\tau} 
+ \mathbf{F}_{g} + \mathbf{F}_{st} 
+ \mathbf{F}_{\text{damp}} + \mathbf{F}_{\text{rec}}.
\label{eq:momentum}
\end{equation}

Where, $\rho$ is the mass density, $P$ is the fluid pressure, and $\boldsymbol{\tau}$ is the viscous stress tensor. In Eq.\ref{eq:momentum}, some additional volumetric source terms are incorporated to account for relevant physics that govern the motion of the molten metal.

$\mathbf{F}_{g}$ denotes the buoyancy force, and it is calculated using the Boussinesq approximation. 

\begin{equation}
\mathbf{F}_g = \rho \mathbf{g} \beta (T - T_{ref}),
\end{equation}

\noindent where $\beta$ is the thermal expansion coefficient and $T_{\text{ref}}$ is the reference temperature. The surface tension force and Marangoni convection are accounted for by the term $\mathbf{F}_{st}$, which is expressed as,

\begin{equation}
    \mathbf{F}_{st} = \left( \sigma \kappa \mathbf{n} + \frac{d\sigma}{dT} \left[\nabla T - \mathbf{n}(\mathbf{n} \cdot \nabla T)\right] \right) |\nabla \varphi_m| \frac{2\rho}{\rho_m + \rho_g}.
\end{equation}

\noindent where, $\sigma$ is the surface tension coefficient, $\kappa$ is the surface curvature of the metal free surface, $\mathbf{n}$ is the unit surface normal vector, $|\nabla \varphi_m|$ is the magnitude of the metal volume fraction, which converts surface forces to volumetric force and the multiplier term $\frac{2\rho}{\rho_m + \rho_g}$ redistributes the interfacial forces towards the heavier phase(metal). $\frac{d\sigma}{dT}$ denotes the temperature gradient of surface tension. 

To model the solidification/melting process, enthalpy-porosity technique \cite{brentENTHALPYPOROSITYTECHNIQUEMODELING988} is used. This method introduces a volumetric damping source term (also known as Carmen-Kozeny sink term) which is denoted by $\mathbf{F}_{damp}$ in the momentum equation (Eq.\ref{eq:momentum}) and expressed as follows, 
\begin{equation}
    \mathbf{F}_{\text{damp}} = \frac{(1-\epsilon)^2}{(\epsilon^3+ 10^{-12})} A_{\text{mush}} (\mathbf{U}-\mathbf{U}_{p}).
\end{equation}

Here, $\epsilon$ is the cell liquid fraction, equal to $1$ in solid cells, $0$ in liquid cells, and between $0$ and $1$ in the mushy zone. $A_{\text{mush}}$ is the mushy zone constant, which is taken as $10^{6}\ \mathrm{kg\,m^{-3}\,s^{-1}}$ \cite{alphonsoPossibilityDoingReduced2023}. $\mathbf{U}_{p}$ is the solid velocity due to the pulling of solidified material out of the domain.

Due to intense volumetric laser heating, the material undergoes vaporization, which generates a force—commonly referred to as recoil pressure—on the surface of the molten metal. This recoil pressure leads to the formation of a depression, known as a keyhole. Several formulations for recoil pressure have been proposed and validated in the literature \cite{knightEvaporationCylindricalSurface1976, semakRoleRecoilPressure1997, chenCalculationModelEvaporation2001}. In general, for atmospheric conditions, the recoil pressure is 0.5-0.6 times the saturated vapor pressure at the vaporization temperature (at 1 atm), due to the recondensation effect after evaporation. However, as the ambient pressure decreases, it is seen from the vacuum laser welding experiments that, the keyhole depth increases, indicating the increase in recoil pressure. 

This trend can be explained by considering the effect of retro-diffusion (recondensation) of vapor molecules. Under atmospheric pressure, evaporated molecules frequently collide with surrounding gas molecules, which results in part of the vapor being redirected back to the molten surface. This retro-diffusion reduces the net escaping vapor flux and, consequently, lowers the effective recoil pressure. In contrast, under reduced ambient pressure, the mean free path of vapor molecules increases significantly, minimizing collisions with the surrounding medium. As a result, recondensation is suppressed, and a greater fraction of evaporated molecules contributes directly to the recoil force.

Therefore, it is essential to incorporate a retro-diffusion parameter in the recoil pressure model, to account for the recondensation effect.

The volumetric recoil pressure source, $\mathbf{F}_{rec}$ is expressed as, 
\begin{equation}
       \mathbf{F}_{rec} = (P_r - P_{amb})\mathbf{n}|\nabla \varphi_m| \frac{2\rho}{\rho_m + \rho_g}.
\end{equation} 

Where, $P_r$ and $P_{amb}$ are the recoil pressure and ambient pressure respectively. The recoil pressure $P_{r}$ can be expressed as \cite{liNumericalExperimentalStudy2019a}, 

\begin{equation}
P_r =
\begin{cases} 
      \dfrac{1+\beta_R}{2} P_{sat}, & T \geq T_b \\[6pt]
      P_{\text{amb}}, & 0 \leq T < T_b
\end{cases}
\end{equation}
where $\beta_R$ is the retro-diffusion coefficient, which depends on the Mach number at the exit of the Knudsen layer. Its value changes from 0.18 to 1, with a decrease in ambient pressure \cite{knightEvaporationCylindricalSurface1976}. $T_b$ is the evaporating temperature at the corresponding ambient pressure. $P_{sat}$ is the temperature-dependent vapor pressure and is calculated using the Clausius-Clapeyron law \cite{linModernThermodynamicsHeat1999}, which is expressed as, 
\begin{equation}
P_{\text{sat}} = P_{\text{amb}} \, \exp\!\left( \frac{M_l L_{v}}{R} \, \frac{T - T_{b}}{T \, T_b} \right),
\end{equation}

\noindent where, $M_l$ is the molar mass of the vaporized metal, R is the gas constant.

Finally, Eq.~\ref{eq:energy} describes the conservation of energy in the computational domain:

\begin{equation}
\frac{\partial(\rho c_{p}T)}{\partial t} 
+ \nabla \cdot (\mathbf{U}\rho c_{p}T) 
- \nabla \cdot (k \nabla T) 
= Q_{\text{laser}} + S_{h} - Q_{\text{losses}}.
\label{eq:energy}
\end{equation}

Where $C_p$ and $k$ are the specific heat and thermal conductivity of the mixture of two phases(gas and molten metal). The phase change effects during the melting are calculated using the $S_h$ term.

The term $Q_{losses}$ in the energy conservation equation (Eq. \eqref{eq:energy}) represents the heat losses during the laser welding process, which includes the convective ($Q_{conv}$) and radiative ($Q_{rad}$) and Evaporative ($Q_{evap}$) heat loss terms, 

\begin{align}
Q_{\text{losses}} &= \bigl( Q_{\text{conv}} + Q_{\text{rad}} + Q_{\text{evp}} \bigr) 
\left|\nabla \varphi_m\right| 
\left( \frac{2\rho}{\rho_m + \rho_g} \right), \label{eq:Qlosses} \\
Q_{\text{rad}}   &= \sigma_{\text{SB}} \, \varepsilon \left( T^4 - T_{\infty}^4 \right), \\
Q_{\text{conv}}  &= h \left( T - T_{\infty} \right), \\
Q_{\text{evp}}   &= L_v \dot{m}.
\end{align}

Here, $\sigma_{SB}$ is the Stephan-Boltzmann constant (value), $\epsilon$ is the emissivity of the material, $h$ is the convective heat transfer coefficient, $T_\infty$ is the ambient temperature. $\dot{m}$ is the mass transfer rate during evaporation, and is approximated using the Hertz-Langmuir relation \cite{kolasinskiSurfaceScienceFoundations2012}, which can be written as, 
\begin{equation}
\dot{m} =
\begin{cases}
0, & T < T_b \\[2mm]
(1-\beta_R) \, \sqrt{\dfrac{M_l}{2 \pi R T}} \, P_{\text{sat}}, & T \ge T_b
\end{cases}
\end{equation}

The laser heat source model ($Q_{laser}$) in Eq.~\ref{eq:energy} follows a Gaussian distribution that scans through the top surface of the substrate at a constant speed and can be epxressed as, 

\begin{equation}
    Q_{laser} = I_0(r)(\mathbf{I_0 \cdot n_0})\eta(\theta_0) + \sum_{m=1}^{N} I_m(r)(\mathbf{I_m \cdot n_m})\eta(\theta_m)
\end{equation} 

\begin{equation}
I_0(r) = \frac{2P}{\pi r_0^2 \Delta} \, \exp\!\left[ \frac{-2 (r - V_s t)^2}{r_0^2} \right]
\end{equation}

where, $\theta$ is the angle between the incident laser beam and normal vector of the keyhole wall, $\eta_{Fr}$ is the Fresnel absorption coefficient, N is the number of laser beam incidences considering multiple reflections, $\mathbf{I}$ is normalized laser beam direction, $\mathbf{n}$ is normalized normal vector of the keyhole wall. $I$ represents the laser energy intensity. The subscript $0$ refers to the incident beam and $m$ denotes the $m^{th}$ reflections. $P$ is the total deposited beam power, $\Delta$ is the computational cell length. $r_0$ is the beam radius($1/e^2$) and $V_s$ is the laser scanning speed. 

To account for multiple reflection during the laser welding process, a high-fidelity ray-tracing method has been implemented in the solver. When each ray hits a computational cell within the domain, part of it gets absorbed and the rest is reflected. The reflection vector is computed using the following equation, 

\begin{equation}
    \mathbf{V} _r = \mathbf{V}_I - \frac{2V_I \cdot \mathbf{n}}{|\mathbf{n^2}|} \mathbf{n},
\end{equation}

\begin{figure}[H]
    \centering
    \includegraphics[width=\textwidth]{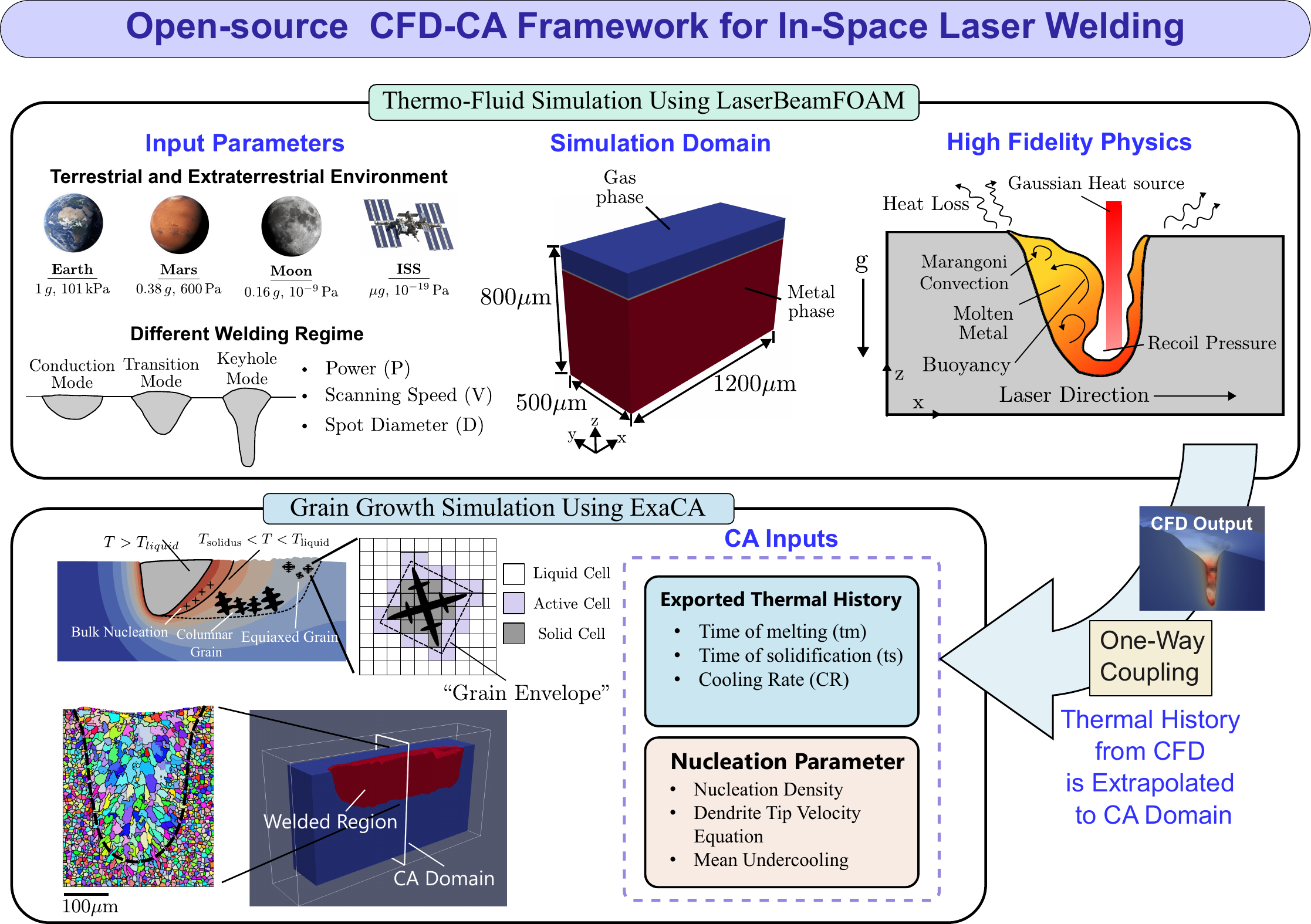}
    \caption{Schematic overview of the open-source one-way coupled CFD-CA framework for simulating laser welding processes in terrestrial and extraterrestrial environments. The computational framework integrates fluid dynamics and solidification modeling to predict meltpool behavior and microstructure evolution under varying gravitational and welding conditions.}
    \label{fig:overview}
\end{figure}

\noindent where, $\mathbf{V}_R$ is the reflected ray vector and $\mathbf{V}_I$ is the incident ray vector. The optical equations used for calculating the laser absorptivity is explained in Supplementary methods 1.

\begin{figure}
    \centering
    \includegraphics[width=\textwidth]{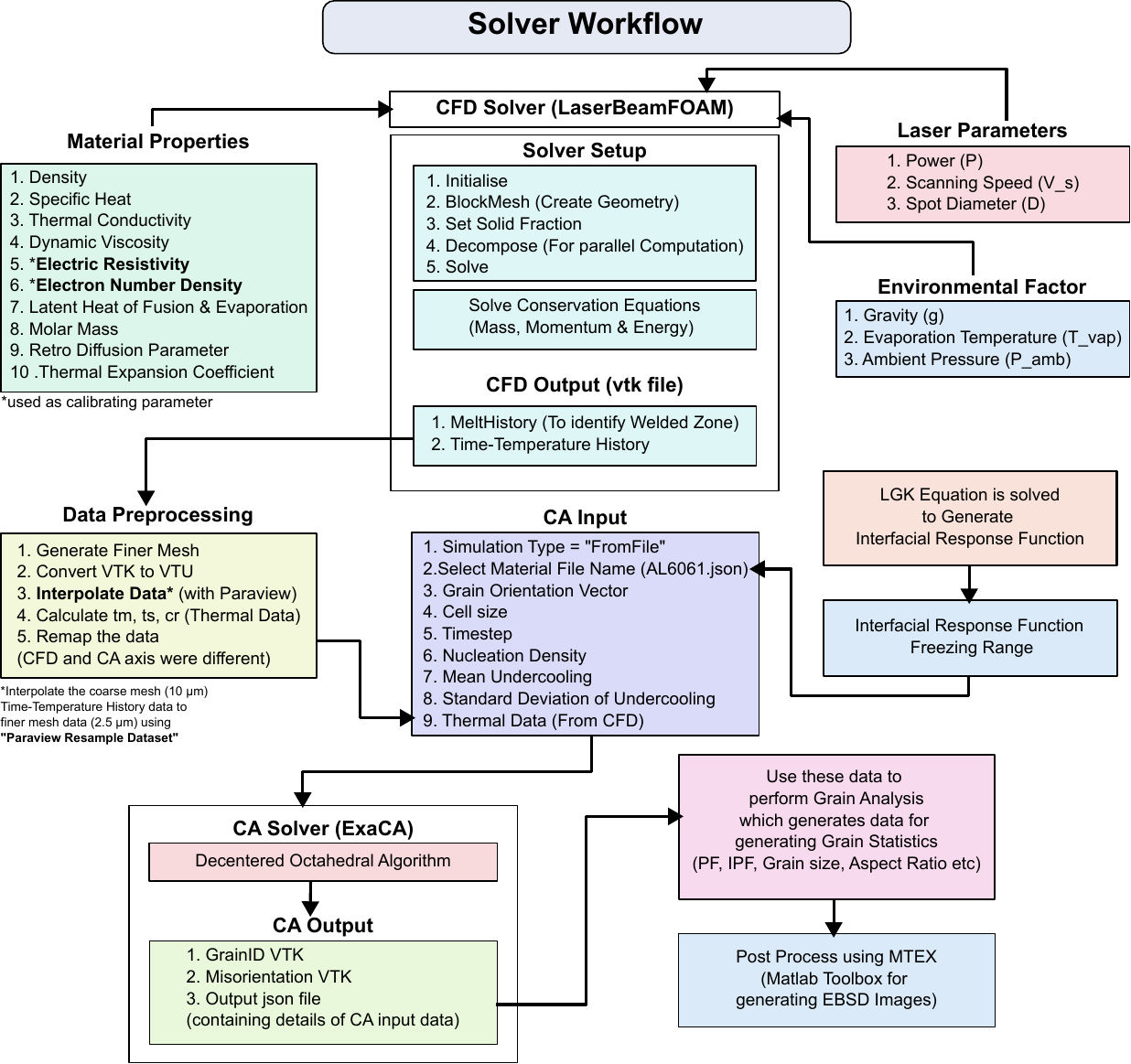}
    \caption{The workflow that connects different modules of the open source software implementation.}
    \label{fig:workflow}
\end{figure}

To capture the gas-liquid interface, the Volume of Fluid (VOF)\cite{leeModelingHeatTransfer2016} method is used. The VOF equation can be written as, 

\begin{equation}
\frac{\partial \varphi}{\partial t} + \nabla \cdot (\mathbf{U} \varphi) = 0,
\end{equation}

\noindent where $V_F$ implies the metal volume fraction within the cell. $V_F$ = 1 implies the cell is filled with the metal phase, whereas $V_F$ = 0 means the cell is entirely filled with the gas phase. Values in between indicate the existence of a free surface.

\subsection{Solver Setup and Geometry}

The computational domain for the thermo-fluid simulation consists of a rectangular region with dimensions of 1200~$\mu$m (length) $\times$ 500~$\mu$m (width) $\times$ 800~$\mu$m (height), discretized using a uniform mesh size of 10~$\mu$m. Simulations were performed with a time step on the order of $1 \times 10^{-7}$~s, and the Courant number was maintained at 0.25 to ensure numerical stability and convergence. The thermal history obtained from the CFD solver was subsequently exported and interpolated onto a finer mesh with 2.5~$\mu$m grid spacing for the CA simulations. The CA solver outputs included grain identification (grain ID) and misorientation angle distributions, which were used to perform grain analysis and extract microstructural characteristics. The complete computational workflow is illustrated in Fig.~\ref{fig:workflow}.

\subsection{Grain Growth Modeling}

ExaCA, an open-source microstructure modeling code \cite{rolchigoExaCAPerformancePortable2022}, is employed to simulate the grain structures. It adopts a Cellular Automata (CA) framework to capture grain-scale microstructure evolution in metallic alloys under rapid solidification conditions. The model considers only the thermal history of the coordinates of the locations where the local temperature exceeds the liquidus temperature during the thermo-fluid simulation. For such locations, the time when each cell reaches the liquidus temperature ($tm$), the time when the cell falls below the solidus temperature ($ts$), and the instantaneous cooling rates ($cr$) were given as thermal history input. The instantaneous cooling rate at each coordinate was calculated from the slope of time-temperature data between the solidus temperature (600°C) and a point 100°C below this threshold, following the NIST AM bench 2018 challenge 02 guideline. \cite{noauthor_amb2018-02_2018} In our case, the CA cell size is four times smaller than the CFD cell size; therefore, the trilinear interpolation is implemented to pair the data with CA.
 
The growth of arbitrarily oriented 3D grains is tracked using the decentered octahedron algorithm \cite{gandinCoupledFiniteElementcellular1994}, which enables grains to assume non-uniform shapes under non-uniform thermal conditions. This algorithm models each grain as an octahedral envelope that expands along the six crystallographic $<001>$ directions. Cells at or below the solidus temperature are marked as solid cells, while those at or above the liquidus temperature are marked as liquid cells. The moving solid-liquid boundary is represented by active cells, each carrying a small octahedral envelope that grows into the surrounding liquid cells. When an envelope impinges on the center of a neighboring liquid cell, that cell is converted into a new active cell and inherits the same grain identity. Active cells that have no liquid neighbors are not tracked. Grain growth proceeds through independent expansion of these envelopes, which are updated each time step according to an interfacial response function (IRF). The IRF is expressed as a cubic polynomial of the local undercooling, which is derived for Al6061 in this study using the Lipton–Glicksman–Kurz (LGK) model \cite{panwisawas_mesoscale_2017}. This equation relates dendritic tip velocity to local undercooling as:
\begin{equation}
\nu (T) = {A} \cdot \Delta {T^3}+  {B} \cdot \Delta {T^2} + {C} \cdot \Delta T +  {D}
\end{equation}

Here, $\Delta T$ is the local undercooling of a cell. The values of the fitting coefficients, $A$, $B$, $C$, $D$ are listed in Table~\ref{tab:exaca_properties}, and the detailed calculations for determining these coefficients are documented in the Supplementary method 2.

Nucleation is modeled heterogeneously. At initialization, a fraction of the liquid cells is selected randomly as potential nuclei. Each potential nucleus is assigned a nucleation undercooling drawn from the Gaussian distribution of undercooling:

\begin{equation}
f(\Delta T) = \frac{1}{\Delta T \sqrt{2\pi}}
\exp\left[
    -\frac{1}{2}
    \left(
        \frac{\Delta T_N - \Delta T}{\Delta T_\sigma}
    \right)^2
\right].
\end{equation}

Here, mean nucleation undercooling is denoted by $\Delta T_N$, and standard deviation is denoted by $\Delta T_\sigma$.

The number of potential nucleation sites in the bulk is calculated as \cite{lianCellularAutomatonFinite2019}:
\begin{equation}
{N_v} = {N_0}  \cdot V
\end{equation}
Here, V denotes the total volume. The values of the input parameters $\Delta T_N$, $\Delta T_\sigma$ and $N_0$ are listed in table~\ref{tab:exaca_properties}.
If a liquid cell designated as a potential nucleus is not captured by an advancing grain and its local undercooling reaches its assigned nucleation undercooling, it becomes an active cell, and a new octahedron with a random orientation is assigned.

\begin{table}[htbp]
  \centering

  \caption{ExaCA Parameters for Microstructure Simulation}
  \label{tab:exaca_properties}
  \begin{tabular*}{\textwidth}{@{\extracolsep{\fill}} l c c c c}
    \toprule
    \textbf{Property} & \textbf{Unit} & \textbf{AL6061} & \textbf{IN625} & \textbf{IN718}\\
    \midrule
    Initial grain size, $S_0$                
        & \si{\micro\meter}                      
        & 8.3
        & 8.3
        & \\
        
    Cell size, $\Delta x$                    
        & \si{\micro\meter}                      
        & 1.25
        & 1.25
        & \\

    3rd order response, $A$ 
        & \si{\meter\per\kelvin\cubed\per\second} 
        & \num{-1.50863e-5}     
        & \num{-1.0302e-7} 
        & \num{2.29e-6} \\

    2nd order response, $B$ 
        & \si{\meter\per\kelvin\squared\per\second} 
        & \num{3.67154e-3}      
        & \num{1.0533e-4} 
        & \num{1.58e-5} \\

    1st order response, $C$ 
        & \si{\meter\per\kelvin\per\second}          
        & \num{1.90518e-2}      
        & \num{2.2196e-3} 
        & \num{1.77e-5} \\

    0th order response, $D$ 
        & \si{\meter\per\second}                     
        & \num{-2.55361e-2}     
        & \num{-2.55361e-2} 
        & \\

    Nucleation density, $N_0$     
        & \si{\per\meter\cubed}                  
        & \num{1e15}
        & \num{1e15}
        & \\

    Mean undercooling, $\Delta T_N$          
        & \si{\kelvin}                           
        & \num{5}
        & \num{5}
        & \\

    Std.\ dev.\ undercooling, $\Delta T_\sigma$ 
        & \si{\kelvin}                           
        & \num{0.5}
        & \num{0.5}
        & \\
                                                      
    \bottomrule
  \end{tabular*}
  
  \vspace{0.5em}
  \raggedright
  \textbf{Note:} (a) Interfacial response modeled as cubic: 
  $V(\Delta T)=A\Delta T^3+B\Delta T^2+C\Delta T+D$.  
  (b) Since IN718 was used only for LGK validation, only its IRF coefficients are presented.
\end{table}

\section{Results \& Discussion}
\subsection{Model Calibration}
To establish the reliability of the coupled thermo-fluid and microstructural models, the simulation results are validated against experimental literature. The validation is conducted in two stages: first, by comparing the transient keyhole dynamics and melt pool characteristics with experimental measurements; and second, by qualitatively and quantitatively evaluating the predicted grain morphology against the experimentally observed microstructure.

\begin{figure}
    \centering
    \includegraphics[width=\textwidth]{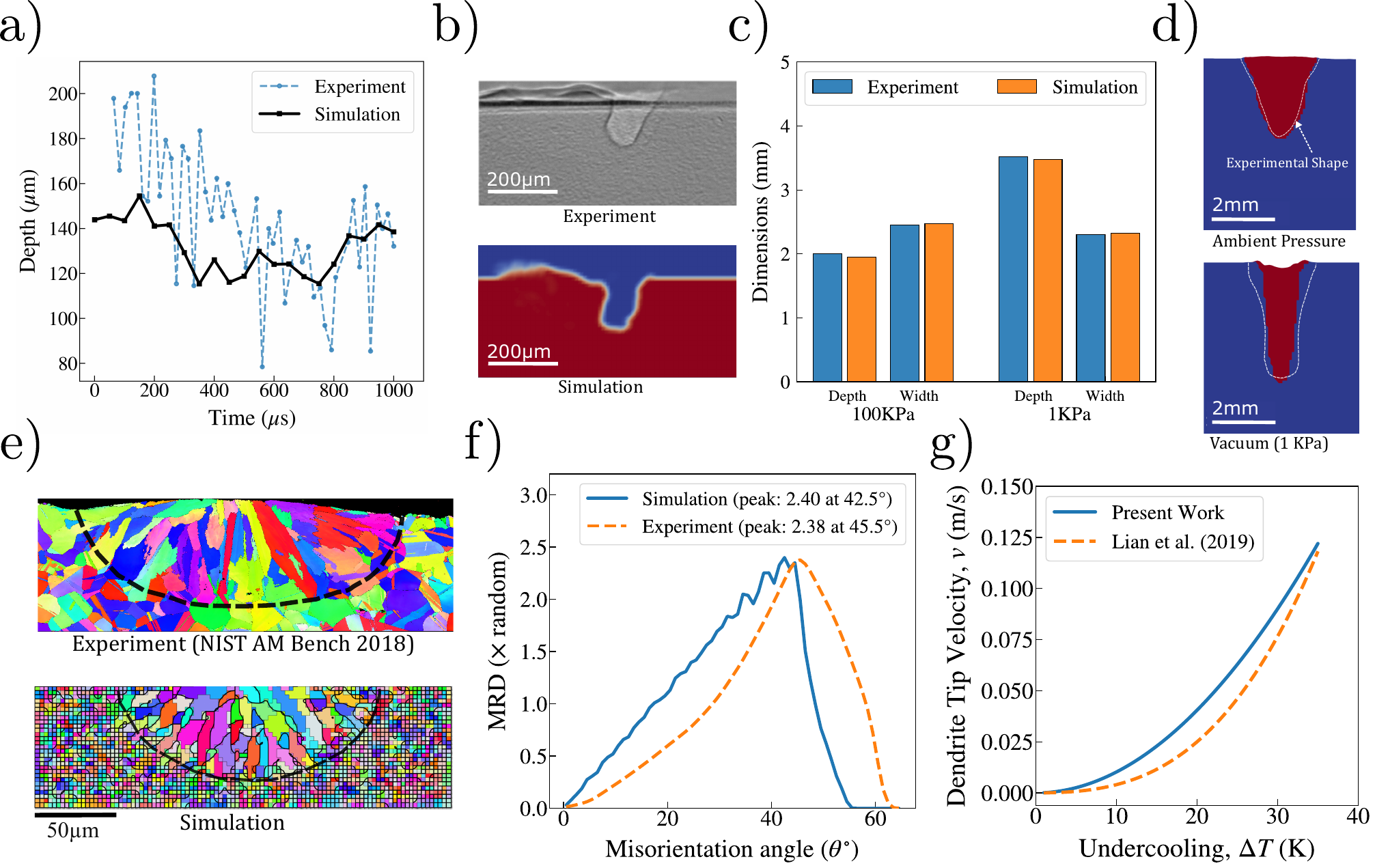}
    \caption{\textbf{Validation of thermo-fluid model (a--d) and microstructure model (e--g) against experimental results.} \textbf{a}, Comparison of predicted keyhole depth during laser welding of $AL6061$ with experimental data reported by Zhang et al. \cite{zhangAccurateEfficientPredictions2025}; process parameters: laser power $P$ = 500 W, scanning velocity $v$ = 0.7 m/s, beam radius $r_0$ = 50 $\mu$m; results shown at $t$ = 1 ms after establishment of quasi-steady keyhole morphology. \textbf{b}, Qualitative comparison of keyhole morphology with experimental X-ray image under identical process parameters. \textbf{c}, Quantitative comparison of melt pool depth and width under atmospheric and vacuum (1 kPa) conditions. \textbf{d}, Melt pool cross-sections comparing the melt pool depth and width with experimental measurements from Lee et al. \cite{leeVacuumLaserBeam2021} for laser welding of $AL6061$; dashed lines indicate experimental melt pool boundaries; process parameters: laser power $P$ = 2100 W, scanning velocity $v$ = 2 m/min, spot diameter $d_0$ = 356 $\mu$m. \textbf{e}, Comparison of simulated grain structure for laser-welded IN625 with experimental EBSD data from NIST AM Bench 2018 Challenge 02; process parameters: laser power $P$ = 137.9 W, scanning velocity $v$ = 0.4 m/s, beam radius $r_0$ = 50 $\mu$m. \textbf{f}, Dendrite tip velocity as a function of undercooling temperature for IN718 predicted using the LGK model, compared with results provided by Lian et al. \cite{lianCellularAutomatonFinite2019}. \textbf{g}, Misorientation angle distribution along the build direction for the NIST AM Bench 2018 case compared with experimental measurements from Yuan et al. \cite{yuanUncoveringGrainSubgrain2024}.}
    \label{fig:validation}
\end{figure}

We first validate the thermo-fluid model against transient keyhole measurements \cite{zhangAccurateEfficientPredictions2025} obtained for a single-track scan on bareplate AL6061 using a 500 W laser at 0.7 m/s with a 50 $\mu$m spot radius. Figure \ref{fig:validation}a compares the predicted and measured keyhole depth as a function of time, while Figure~\ref{fig:validation}b provides a qualitative comparison of the keyhole morphology under the same conditions. Together, these results show that the model reproduces the keyhole dynamics with good fidelity. The thermophysical and laser properties used for AL6061 are presented in Supplementary Table 1.

As this study focuses on laser welding applications in various space environments, validating the model's reliability under both atmospheric and vacuum conditions is essential. To achieve this, single-track bare-plate AL6061 laser welding simulations were performed under both atmospheric and vacuum (1 kPa) conditions, following the experimental process parameters reported by Lee et al. \cite{leeVacuumLaserBeam2021}: laser power of 2100 W, scanning velocity of 2 m/min, and spot diameter of 356 $\mu$m.

Fig.\ref{fig:validation}c and fig.\ref{fig:validation}d illustrates the cross-sectional melt pool morphology under two conditions. Under ambient pressure, the melt pool exhibits a characteristic nail-head profile with a wider surface width and gradual tapering toward the penetration depth. In contrast, under vacuum conditions (1 kPa), the melt pool displays a deeper, narrower morphology with steeper sidewall angles and increased penetration depth. The white contour lines delineate the experimental meltpool shape in each case. This transition from a shallow, wide profile to a deep, narrow profile is attributed to  vapor-induced recoil pressure effects in the low-pressure environment, which enhances keyhole formation and increases penetration depth.

The second phase of the validation process addresses the microstructure model. For this purpose, the NIST Additive Manufacturing Benchmark 2018 (AMB2018-02) test case \cite{laneMeasurementsMeltPool2020} was employed, in which a single-track laser weld was deposited on a bare IN625 substrate using a laser power of 137.9 W, scan speed of 0.4 m/s, and beam radius of 50 $\mu$m. The thermophysical properties, laser settings and CA properties for IN625 are provided in the supplementary methods. A qualitative comparison of the Inverse Pole figure (IPF-Z) of both simulation and experiment is presented in Fig.~\ref{fig:validation}e. 

To further validate the model quantitatively, we compared the grain boundary misorientation distribution along the build direction (z-axis) for this meltpool cross-section. The Multiples of Random Distribution (MRD), which represents the relative frequency of grain boundary misorientations normalized to a random distribution, was calculated and compared between simulation and experiment as presented in Fig.~\ref{fig:validation}f, where the simulated results show good agreement with experimental measurements.

Accurate prediction of grain formation kinetics in the CA model requires establishing the relationship between dendrite tip velocity and undercooling temperature. This relationship is typically derived using LGK/KGT models and has been reported for IN718 \cite{lianCellularAutomatonFinite2019}; however, corresponding data for AL6061 were not directly available in the literature. To validate our implementation, we first solved the LGK equations for IN718 and compared the results with published data \cite{lianCellularAutomatonFinite2019}, as shown in Fig.~\ref{fig:validation}f. We then applied the same methodology to AL6061, treating it as a pseudo-binary alloy for computational simplicity. The resulting dendrite tip velocity--undercooling relationship for AL6061 is presented in Fig.~\ref{fig:dendrite_tip}. The detailed calculation procedure and required thermophysical properties are provided in the supplementary methods section.

\subsection{Parametric Study Design: Laser Welding under different space environments}

Following validation of the coupled thermo-fluid and microstructural models, a systematic parametric investigation was conducted to elucidate the influence of gravitational and atmospheric conditions on laser welding behavior of AL6061. The study encompasses 3 distinct welding modes—conduction, transition, and keyhole—across 4 representative space environments: Earth (baseline), Mars, Lunar, and ISS (International Space Station) microgravity conditions. The substrate initial temperature ($T_0$) was assumed to be constant at 300 K for all cases. 

\begin{table}[htbp]
  \centering
  \caption{Summary of simulation and validation cases for laser welding under different gravitational environments.}
  \label{tab:case_matrix}
  \small
  \begin{tabular}{@{}llcccccc@{}}
    \toprule
    \textbf{Case} & \textbf{Environment} & \textbf{Gravity, g} & \textbf{Pressure, $P_{\text{amb}}$} & \textbf{$T_{\text{vap}}$} & \textbf{Power, P} & \textbf{Speed, V} & \textbf{$E_v$} \\
    & & \textbf{(m/s$^2$)} & \textbf{(Pa)} & \textbf{(K)} & \textbf{(W)} & \textbf{(m/s)} & \textbf{(J/mm$^3$)} \\
    \midrule
    \multicolumn{8}{c}{\textbf{Simulation Cases}} \\
    \midrule
    1 & Earth & 9.81 & 101325 & 2792 & 300 & 0.75 & 57.7 \\
    2 & (Baseline) & & & & 370 & 0.5 & 106.8 \\
    3 & & & & & 520 & 0.45 & 166.8 \\
    \midrule
    4 & Mars & 3.73 & 600 & 1967.86 & 300 & 0.75 & 57.7 \\
    5 & & & & & 370 & 0.5 & 106.8 \\
    6 & & & & & 520 & 0.45 & 166.8 \\
    \midrule
    7 & Lunar & 1.67 & $1\times10^{-4}$ & 1036.71 & 300 & 0.75 & 57.7 \\
    8 & & & & & 370 & 0.5 & 106.8 \\
    9 & & & & & 520 & 0.45 & 166.8 \\
    \midrule
    10 & Microgravity & $9.81\times10^{-6}$ & $1\times10^{-4}$ & 1036.71 & 300 & 0.75 & 57.7 \\
    11 & & & & & 370 & 0.5 & 106.8 \\
    12 & & & & & 520 & 0.45 & 166.8 \\
    \midrule
    \multicolumn{8}{c}{\textbf{Validation Cases}} \\
    \midrule
    V1 & Earth & 9.81 & 101325 & 2792 & 500 & 0.7 & 91.1 \\
    V2 & Earth & 9.81 & 101325 & 2792 & 2100 & 0.0333 & 631.9 \\
    \bottomrule
  \end{tabular}

  \begin{tablenotes}
    \small
    \item Note: Beam radius $r_0$ = 47 $\mu$m for simulation cases. For validation cases: V1 has $r_0$ = 50 $\mu$m, V2 has $r_0$ = 178 $\mu$m. Volumetric energy density $E_v = P/(V \cdot \pi r_0^2)$. Actual ambient pressures: Earth = 101325 Pa, Mars = 600 Pa, Lunar = $1\times10^{-9}$ Pa, Microgravity = $1\times10^{-19}$ Pa. Simulation pressure for vacuum environments was set to $1\times10^{-4}$ Pa for numerical stability and tackling sublimation modeling issue.
  \end{tablenotes}
\end{table}

Table~\ref{tab:case_matrix} summarizes the 12 simulation cases investigated in this study. Each environment is characterized by its gravitational acceleration and ambient pressure, which directly influence the vapor pressure equilibrium and keyhole stability. For Mars surface conditions, an ambient pressure of 600 Pa was applied, corresponding to the typical Martian atmospheric pressure. Lunar and microgravity environments were simulated under high vacuum conditions (1×10$^{-4}$ Pa simulation pressure), representative of the ultra-low pressure space environment. The corresponding evaporation temperatures ($T_{\text{vap}}$) were calculated based on the Clausius-Clapeyron relation for each pressure condition.

To study the different welding regime across different space environments, three laser power–velocity combinations were selected : 300~W at 0.75~m/s, 370~W at 0.7~m/s, and 520~W at 0.45~m/s. The beam radius was held constant at 47~$\mu$m for all cases. These parameter combinations yield volumetric energy densities ($E_v$) of 57.7, 106.8, and 166.8~J/mm$^3$, respectively, representing distinct operating regimes spanning from shallow conduction-limited melting to deep keyhole penetration. This matrix enables direct comparison of how identical process parameters perform under different space conditions, as well as how different welding modes respond to varying gravity and pressure environments.


\subsection{Prediction of meltpool geometry}

\begin{figure}
    \centering
    \includegraphics[width=\textwidth]{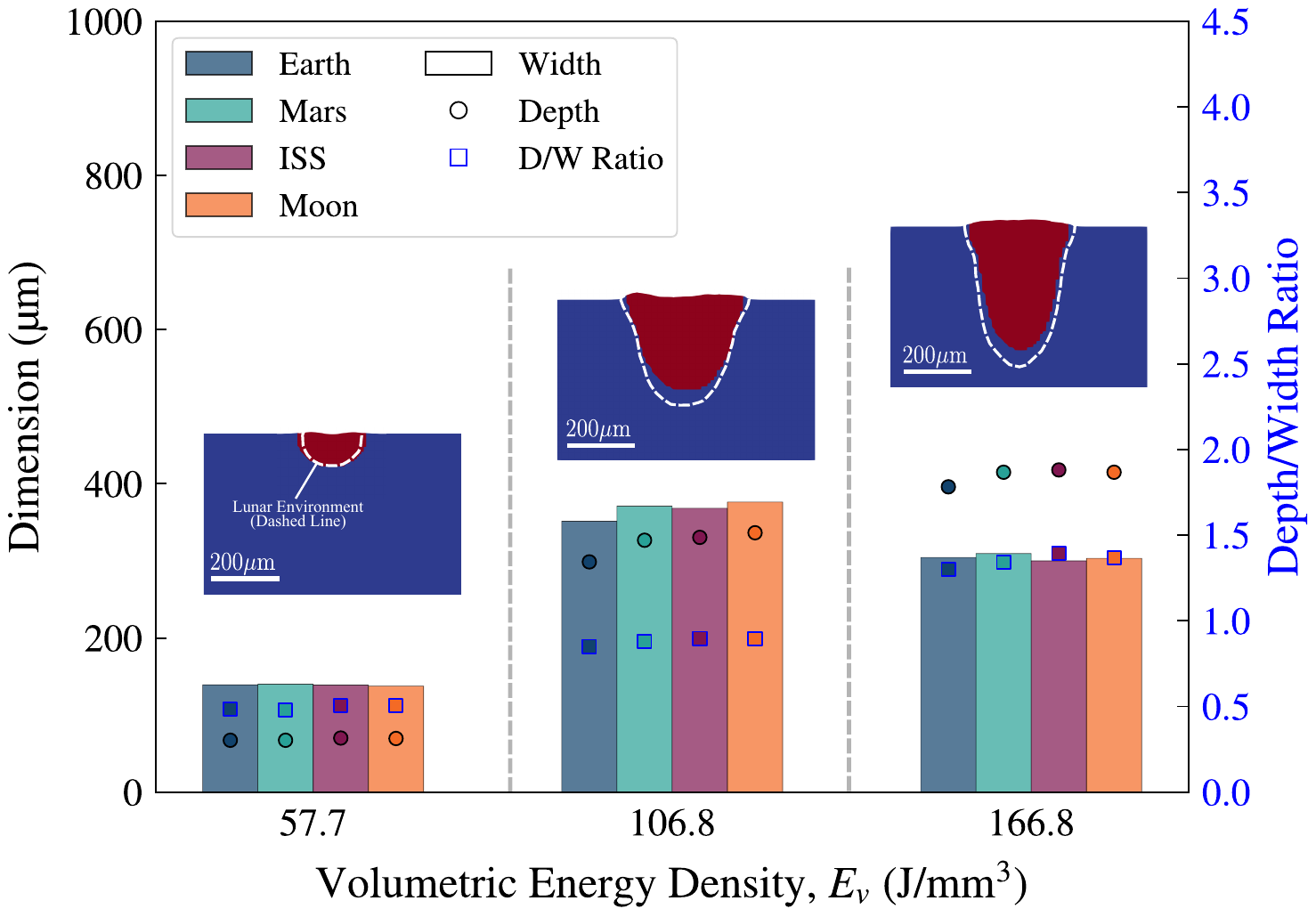}
    \caption{Meltpool characteristics as a function of volumetric energy density ($E_v$ = 57.7, 106.8, and 166.8~J/mm$^3$) across four gravitational environments. Bars represent meltpool width ($\mu$m, left axis), circles indicate depth ($\mu$m, left axis), and squares show the depth-to-width ratio (right axis) for Earth, Mars, ISS, and Moon conditions. Meltpool morphology for each case is shown above the corresponding bar group, where the red region depicts the meltpool shape under Earth conditions and the white dashed line indicates the meltpool shape under lunar conditions.}
    \label{fig:meltpool_result}
\end{figure}

The predicted meltpool depth, width, depth-to-width ratio and meltpool shapes for the three $E_v$ across the four space environments is presented in Fig.~\ref{fig:meltpool_result}. The depth-to-width ratio serves as an initial indicator of welding mode, where $2d/w<1$ suggests conduction-dominated melting and $2d/w>1$ indicates keyhole formation \cite{laneTransientLaserEnergy2020}. At the lowest energy density (57.7~J/mm$^3$), the $2d/w$ ratio ranges from 0.97 to 1.01 across all environments, clearly indicating conduction mode behavior. For the medium (106.8~J/mm$^3$) and high (166.8~J/mm$^3$) energy density cases, the $2d/w$ ratios are 1.70--1.79 and 2.60--2.74 respectively, both exceeding unity and suggesting keyhole mode. 

However, the keyhole number ($K_e$) \cite{ganUniversalScalingLaws2021} provides a more precise classification of welding modes. The keyhole number is defined as:
\begin{equation}
K_e = \frac{\eta P}{(T_l - T_0) \pi \rho c_p \sqrt{\alpha V r_0^3}}
\end{equation}
where $\eta$ is the laser absorptivity, $P$ is the laser power, $T_l$ is the liquidus temperature, $T_0$ is the initial substrate temperature, $\rho$ is the density, $c_p$ is the specific heat capacity, $\alpha$ is the thermal diffusivity, $V$ is the scanning velocity, and $r_0$ is the beam radius. The welding mode is classified based on the keyhole number as:
\begin{equation}
\text{Welding Mode} = 
\begin{cases}
\text{Conduction,} & K_e < 1.4 \\
\text{Transition,} & 1.4 \leq K_e \leq 6 \\
\text{Keyhole,} & K_e > 6
\end{cases}
\end{equation}
Based on this criterion, $K_e = 4.42$ for the medium energy density case, classifying it as transition mode ($1.4 < K_e < 6$), while $K_e = 10.09$ for the high energy density case indicates keyhole mode ($K_e > 6$). Therefore, the three parameter combinations represent conduction, transition, and keyhole welding modes respectively.

Comparing Earth with lunar conditions reveals distinct mode-dependent sensitivities. In conduction mode (57.7~J/mm$^3$), the Moon shows minimal deviation from Earth with only 1.2\% narrower width (137.6 vs 139.2~$\mu$m) and 3.4\% greater depth (69.6 vs 67.3~$\mu$m). Transition mode (106.8~J/mm$^3$) exhibits significantly higher environmental sensitivity, where lunar welding produces 7.1\% wider pools (376.3 vs 351.4~$\mu$m) and 12.7\% deeper penetration (336.5 vs 298.7~$\mu$m) compared to Earth. This enhanced sensitivity reflects the altered melt pool dynamics in reduced gravity, where convective flows and surface tension-driven phenomena play more prominent roles. In keyhole mode (166.8~J/mm$^3$), the Moon again shows modest increases of 0.4\% in width (303.3 vs 304.5~$\mu$m) and 4.8\% in depth (415.0 vs 396.1~$\mu$m), indicating that keyhole dynamics are relatively robust against gravitational variations once deep penetration is established.

While extraterrestrial conditions consistently differ from Earth, the variations among the three space environments themselves (Mars, ISS, and Moon) are remarkably small. In conduction mode, the maximum difference in width among these three environments is only 1.2~$\mu$m (138.9--140.1~$\mu$m, 0.9\% relative to the mean), and depth varies by 2.9~$\mu$m (67.3--70.2~$\mu$m, 4.3\% relative to the mean). For transition mode, the inter-space spread is 8.0~$\mu$m for width (368.3--376.3~$\mu$m, 2.2\% of mean) and 9.7~$\mu$m for depth (326.8--336.5~$\mu$m, 2.9\% of mean). Even in keyhole mode, the variation among extraterrestrial cases remains modest at 9.5~$\mu$m for width (300.0--309.5~$\mu$m, 3.1\% of mean) and 3.0~$\mu$m for depth (415.0--418.0~$\mu$m, 0.7\% of mean).

The magnitude of Earth-to-space differences observed here is surprisingly small compared to what is typically reported in the literature. This is particularly noteworthy for deep penetration welding, where meltpool depth is usually expected to increase significantly in reduced gravity environments \cite{katayamaDevelopmentDeepPenetration2011}. These relatively modest variations warrant further investigation to understand the underlying physical mechanisms. Nevertheless, the findings suggest that a single optimized parameter set could potentially serve across multiple extraterrestrial destinations with minimal modification, at least within this micro-scale process window—a practical advantage for space manufacturing where power constraints favor lower-power, smaller-beam welding configurations.

\subsection{Effect of length-scale on meltpool and keyhole morphology}
According to the established hypothesis, meltpool depth should increase significantly under vacuum conditions due to enhanced recoil pressure resulting from reduced evaporation temperature. However, Fig.~\ref{fig:meltpool_result} shows that meltpool shape and dimensions remain relatively unchanged compared to terrestrial conditions for the selected process parameters. This discrepancy between theoretical expectations and observed results suggests that the effect of length scale on meltpool dimensions requires further investigation, particularly for transition and keyhole mode welding. The characteristic length scales of the laser beam ($r_0 = 47$~$\mu$m) and resulting meltpool features may play a crucial role in determining the relative importance of gravity-driven convection versus surface tension and recoil pressure-driven flows. 


The pressure needed to keep the keyhole open is balanced by closing terms. The local balance at the keyhole front is \cite{fabbroAnalysisPhysicalProcesses2016},
\begin{equation}\label{eq:keyhole_pressure_balance}
P_0 \;=\; P_{\mathrm{amb}} \;+\; P_c \;+\; P_h \;+\; P(V),
\end{equation}
where \(P_{\mathrm{amb}}\) is ambient (chamber) pressure, \(P_c=\sigma/r\) is the capillary (surface tension) closing pressure with surface tension \(\sigma\) and local keyhole/beam radius \(r\), \(P_h\) is hydrostatic head (usually small), and \(P(V)\) is a speed-dependent contribution that grows with welding speed.

\vspace{0.4em}

When travel speed is low, \(P(V)\approx 0\) and \(P_h\) is negligible compared to \(P_c\), so eq.\ref{eq:keyhole_pressure_balance} reduces to
\begin{equation}\label{eq:balance_low}
P_0 \;\approx\; P_{\mathrm{amb}} \;+\; P_c .
\end{equation}

Reducing the ambient pressure, \(P_{\mathrm{amb}}\), lowers the evaporation temperature, \(T_{\mathrm{vap}}\), thereby decreasing the recoil pressure required to sustain the keyhole. However, once the ambient pressure drops below a certain threshold, the capillary (surface tension) forces begin to dominate. Beyond this regime, further reduction in \(P_{\mathrm{amb}}\) produces little additional effect, and the penetration depth tends to \emph{saturate}. According to Fabbro \emph{et al.}, the corresponding threshold—often referred to as the “critical” ambient pressure—can be estimated as
\begin{equation}\label{eq:Pcr}
    P_{\mathrm{cr}} \approx 0.1\,P_c = 0.1\,\frac{\sigma}{r},
\end{equation}
where \(\sigma\) denotes the surface tension and \(r\) is the characteristic radius of curvature of the melt pool interface.

\vspace{0.4em}

At higher speeds, the dynamic term \(P(V)\) is non-negligible; then the effective closing side becomes
\begin{equation}\label{eq:balance_high}
P_0 \;\approx\; \big(P_{\mathrm{amb}} + P(V)\big) \;+\; P_c ,
\end{equation}
so decreasing \(P_{\mathrm{amb}}\) alone has limited leverage if \(P(V)+P_c\) dominates.

To illustrate how beam radius and welding speed govern the balance in Eq.~\ref{eq:keyhole_pressure_balance}, and how variation in length scale affects the melt-pool depth, we examine two representative cases for Al6061 aluminum alloy ($\sigma \approx 0.9~\mathrm{N\,m^{-1}}$) across different length scales. Process parameters and relevant calculation results are summarized in Table~\ref{tab:cases}. Case V2 from Table~\ref{tab:case_matrix} represents a large beam radius $r = 178~\mu\mathrm{m}$ operating at low speed $V = 0.0333~\mathrm{m\,s^{-1}}$ with power $P = 2100~\mathrm{W}$, yielding a capillary pressure $P_{c} = \sigma/r \approx 5.06~\mathrm{kPa}$ and a critical ambient pressure $P_{\mathrm{cr}} = 0.1\,P_{c} \approx 0.506~\mathrm{kPa}$. At this low speed, $P(V) \approx 0$ and Eq.~\ref{eq:balance_low} applies; reducing $P_{\mathrm{amb}}$ from atmospheric toward $0.5~\mathrm{kPa}$ directly lowers the closing pressure and increases penetration. This analysis is supported by the simulation and experimental results in Fig.~\ref{fig:validation}d, where melt-pool depth is significantly higher at $1~\mathrm{kPa}$ (vacuum), compared to the atmospheric case. However, reducing ambient pressure below this threshold provides diminishing returns as the capillary term dominates.

In contrast, Case 2 from Table~\ref{tab:case_matrix} employs a smaller radius $r = 47~\mu\mathrm{m}$ at high speed $V = 0.5~\mathrm{m\,s^{-1}}$ with power $P = 370~\mathrm{W}$, yielding $P_{c} \approx 19.15~\mathrm{kPa}$ and $P_{\mathrm{cr}} \approx 1.915~\mathrm{kPa}$. One might therefore expect that reducing pressure from $101~\mathrm{kPa}$ to $1.915~\mathrm{kPa}$ would greatly increase penetration depth; however, Fig.~\ref{fig:meltpool_result}b shows that the melt-pool depth increases only modestly in extraterrestrial cases compared to Earth. The main reason is that at this smaller scale the laser scan speed is very high, making the $P(V)$ term in Eq.~\ref{eq:keyhole_pressure_balance} non-negligible; hence Eq.~\ref{eq:balance_high} governs, and reducing $P_{\mathrm{amb}}$ below a few kPa produces little additional penetration unless the speed is simultaneously reduced.

\begin{figure}
    \centering
    \includegraphics[width=\textwidth]{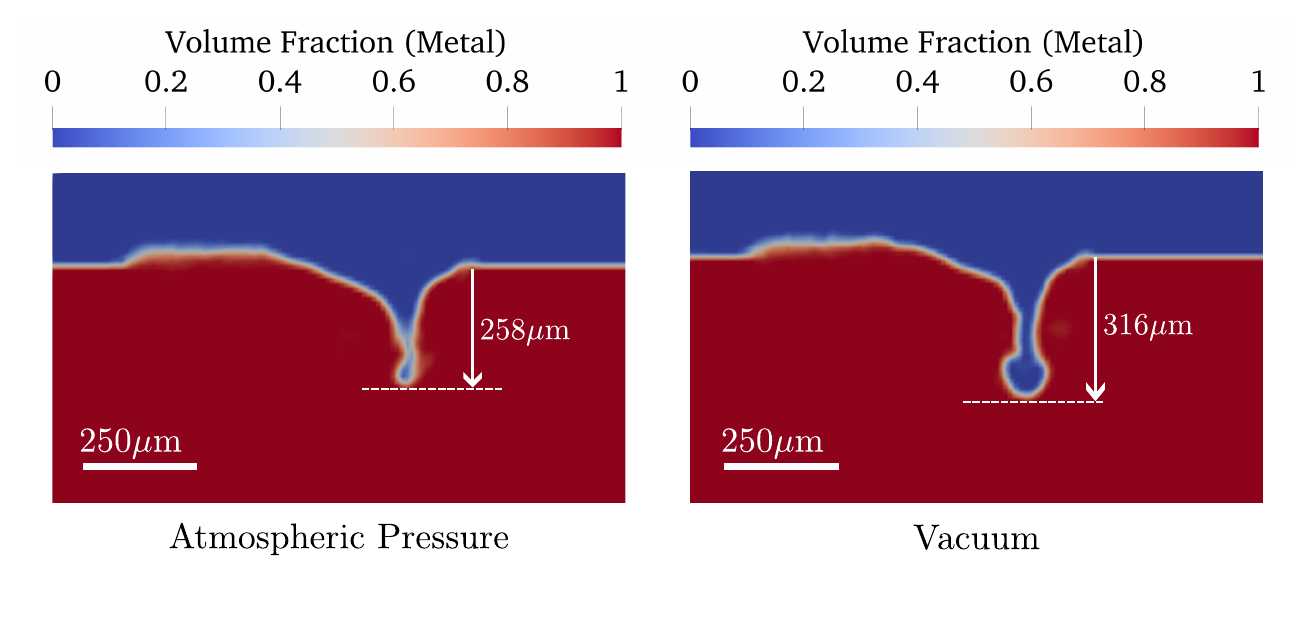}
    \caption{Comparison of keyhole depth between Earth atmospheric conditions and vacuum conditions at reduced scanning speed ($P = 250~\mathrm{W}$, $V = 0.15~\mathrm{m\,s^{-1}}$, $r = 50~\mu\mathrm{m}$), demonstrating the emergence of pressure-driven penetration enhancement when velocity contributions are minimized.}
    \label{fig:hypothesis_proof}
\end{figure}

To verify this hypothesis, additional simulations were conducted at reduced scanning speed: $P = 250~\mathrm{W}$, $V = 0.15~\mathrm{m\,s^{-1}}$, and $r = 50~\mu\mathrm{m}$ for both Earth atmospheric conditions and vacuum conditions ($1 \times 10^{-4}~\mathrm{Pa}$, microgravity). As shown in Fig.~\ref{fig:hypothesis_proof}, the keyhole depth for the atmospheric case is $258~\mu\mathrm{m}$, while for vacuum it increases to $316~\mu\mathrm{m}$—approximately a 22\% increase. Although the slow scanning speed prevents the simulation from reaching fully steady-state keyhole depth within the computational timeframe, this significant difference clearly demonstrates that reducing scan speed at micro-scale indeed enables the expected pressure-driven penetration enhancement in vacuum conditions. The result confirms that the velocity-dependent pressure term $P(V)$ suppresses the ambient pressure effect at high speeds, and that lowering $V$ shifts the system toward the pressure-dominated regime where $P_{\mathrm{amb}}$ plays a critical role.

This observation underscores a critical gap in the current understanding: a systematic investigation is needed to determine the precise parameter window where velocity contributions become negligible and pressure dependence dominates the keyhole dynamics. The transition from the velocity-dominated regime (Eq.~\ref{eq:balance_high}) to the pressure-dominated regime (Eq.~\ref{eq:balance_low}) is not sharply defined but depends on the interplay between beam radius, power, scan speed, and material properties. This distinction has profound implications for space manufacturing strategy. At larger scales ($r > 200~\mu\mathrm{m}$), meltpool dimensions vary significantly across different atmospheric conditions, necessitating environment-specific parameter optimization for each extraterrestrial destination. However, at smaller scales ($r < 50~\mu\mathrm{m}$), the present results demonstrate that environmental variations produce minimal changes in meltpool geometry—a finding with important practical advantages.

The relative insensitivity of micro-scale welding to ambient conditions is particularly beneficial for space applications, where a single optimized parameter set could potentially serve across multiple extraterrestrial environments without substantial modification. Moreover, space-based manufacturing faces severe power constraints, making kilowatt-level laser systems impractical for orbital or planetary operations. Achieving comparable weld quality with lower power inputs is therefore a critical design objective. This research specifically targets the micro-scale regime to explore whether reduced beam size and lower power can deliver consistent welding performance across varying gravity and pressure environments, thereby enabling power-efficient, universally applicable welding protocols for space applications.

Future work should systematically vary scan speed at fixed power and radius in both terrestrial and reduced-gravity environments to map the transition boundaries and establish predictive scaling laws for the relative importance of $P(V)$ versus $P_{\mathrm{amb}}$ in determining keyhole stability and penetration depth. Such systematic mapping will enable the identification of optimal process windows that maximize power efficiency while minimizing sensitivity to environmental variations—key requirements for robust space manufacturing systems.

\begin{table}[h]
\centering
\small
\caption{Process parameters and calculated pressures for two welding regimes in Al 6061 (referenced from Table~\ref{tab:case_matrix}).}
\label{tab:cases}
\begin{tabular}{lcc}
\hline
\textbf{Parameter} & \textbf{Case V2} & \textbf{Case 2} \\
 & \textbf{(Large beam, low speed)} & \textbf{(Small beam, high speed)} \\
\hline
Beam radius, $r$ ($\mu$m) & 178 & 47 \\
Welding speed, $V$ (m/s) & 0.0333 & 0.5 \\
Laser power, $P$ (W) & 2100 & 370 \\
Penetration depth, $h$ (mm) & 1.3 & 0.35 \\
\hline
Capillary pressure, $P_c$ (kPa) & 5.06 & 19.15 \\
Critical ambient pressure, $P_{\mathrm{cr}}$ (kPa) & 0.506 & 1.915 \\
Bond number, $Bo$ & $8.3 \times 10^{-4}$ & $5.1 \times 10^{-5}$ \\
$P_h/P_c$ ratio (\%) & 0.61 & 0.041 \\
\hline
\end{tabular}
\end{table}

\subsection{Effect of gravity on meltpool morphology}

To assess the role of gravity in keyhole stability across different extraterrestrial environments, we examine the dimensionless Bond number \cite{liDynamicsViscousEntrapped2018_bondnumber}, which quantifies the relative importance of gravitational versus capillary forces:
\begin{equation}
Bo \;=\; \frac{\rho g L^2}{\sigma},
\end{equation}
where $L$ is a characteristic length scale. For keyhole welding, we take $L = r$ (the beam radius), giving:
\begin{equation}
Bo \;=\; \frac{\rho g r^2}{\sigma}.
\end{equation}
For a keyhole of depth $h$ and radius $r$, the hydrostatic pressure at the bottom is $P_h = \rho g h$, while the capillary pressure is $P_c = \sigma/r$. The ratio of hydrostatic head to capillary pressure is therefore:
\begin{equation}
\frac{P_h}{P_c} \;=\; \frac{\rho g h}{\sigma/r} \;=\; \frac{\rho g h \cdot r}{\sigma} \;=\; \frac{\rho g r^2}{\sigma} \cdot \frac{h}{r} \;=\; Bo \cdot \frac{h}{r}.
\end{equation}

This ratio determines whether gravitational effects significantly influence the pressure balance in Eq.~\ref{eq:keyhole_pressure_balance}. Using material properties for molten Al 6061 ($\rho \simeq 2400$ kg/m$^3$, $\sigma \approx 0.9$ N/m), we evaluate this ratio for the two representative cases from Table~\ref{tab:case_matrix}. 

For Case V2 operating under Earth gravity ($g = 9.81$ m/s$^2$) with beam radius $r = 178$ $\mu$m and penetration depth $h = 1.3$ mm, the Bond number is $Bo \approx 8.3 \times 10^{-4}$. With an aspect ratio $h/r \approx 7.30$, we obtain $P_h/P_c \approx 0.61\%$, indicating that hydrostatic pressure contributes less than 1\% to the total closing pressure. For Case 2 with smaller beam radius $r = 47$ $\mu$m and penetration depth $h = 350$ $\mu$m, the Bond number decreases to $Bo \approx 5.8 \times 10^{-5}$, and with aspect ratio $h/r \approx 7.45$, the ratio becomes $P_h/P_c \approx 0.043\%$—nearly an order of magnitude smaller than Case V2.

Since $Bo \ll 1$ in both cases under terrestrial gravity, the hydrostatic term $P_h$ remains negligible compared to capillary pressure $P_c$, validating the simplified pressure balances in Eqs.~\ref{eq:balance_low} and \ref{eq:balance_high}. More importantly, this analysis reveals that gravitational effects on keyhole stability are virtually insignificant even at Earth's gravity level. For the reduced-gravity environments considered in this study—Mars ($g = 3.73$ m/s$^2$), Lunar ($g = 1.67$ m/s$^2$), and microgravity ($g \approx 10^{-5}$ m/s$^2$)—the Bond number becomes proportionally smaller, making gravitational contributions to the pressure balance even more negligible.


\subsection{Microstructure Analysis}

A total of 12 simulations were conducted in order to analyze the effect of gravity and Volumetric Energy Density ($E_v$) on grain growth evolution during laser welding of AL6061. An average grain size of 8 $\mu$m was taken to initialize the randomly oriented equiaxed grains in the CA domain. 

\begin{figure}
    \centering
    \includegraphics[width=0.7\textwidth]{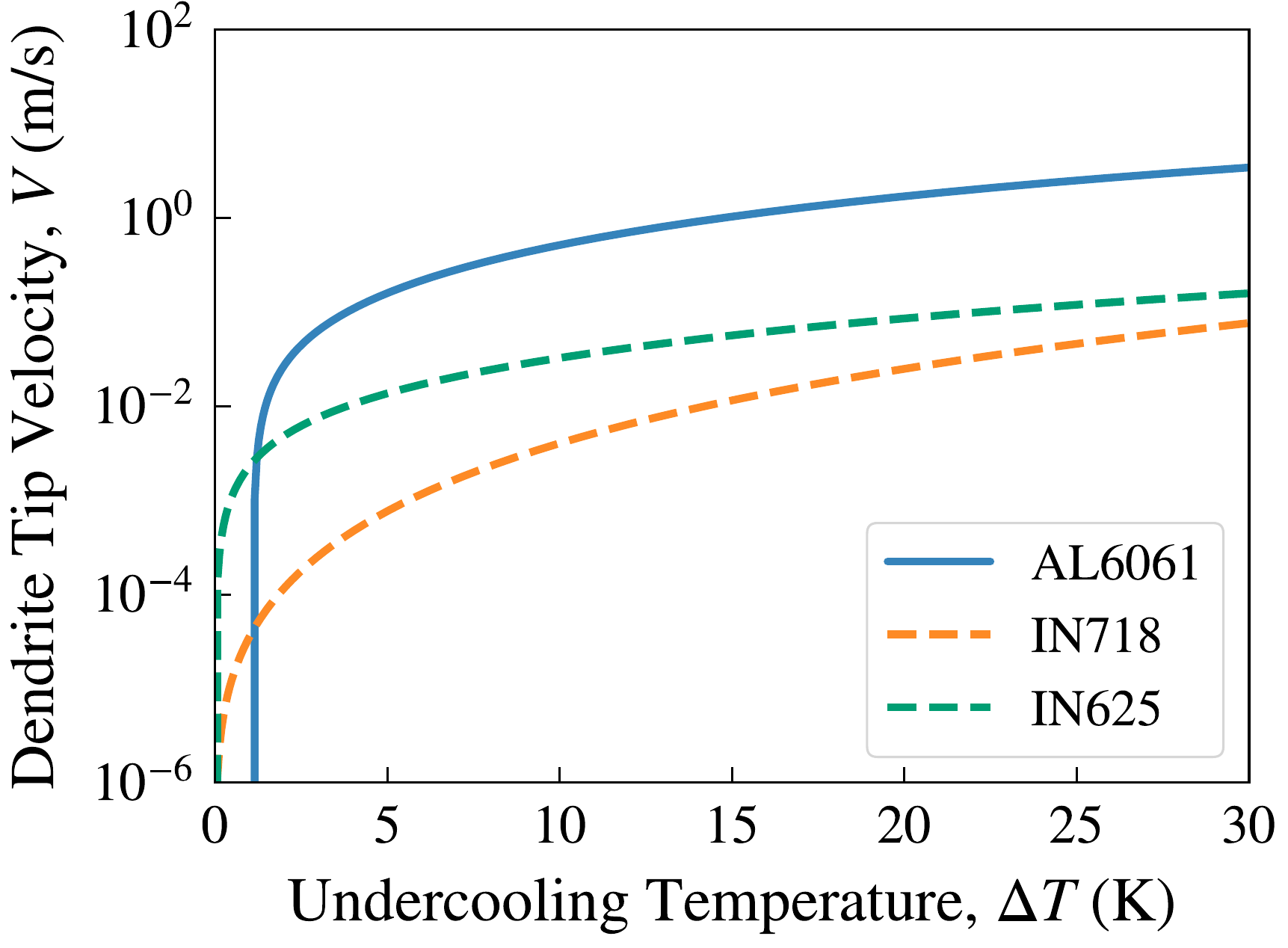}
    \caption{Dendrite tip velocity as a function of undercooling temperature for Al6061, IN718, and IN625. The curves for Al6061 and IN718 are obtained by numerically solving the LGK equations, whereas the IN625 curve is adopted from the literature \cite{rolchigoExaCAPerformancePortable2022}.}
    \label{fig:dendrite_tip}
\end{figure}

Grain growth is typically a result of epitaxial crystallization at the fusion boundary line, which results in subsequent grain growth process. During AM process, the local temperature gradients tend to be high, along with narrow undercooling  region \cite{Shi2025}. Columnar grain growth is usually promoted due to these conditions.

\begin{figure}
    \centering
    \includegraphics[width=\textwidth]{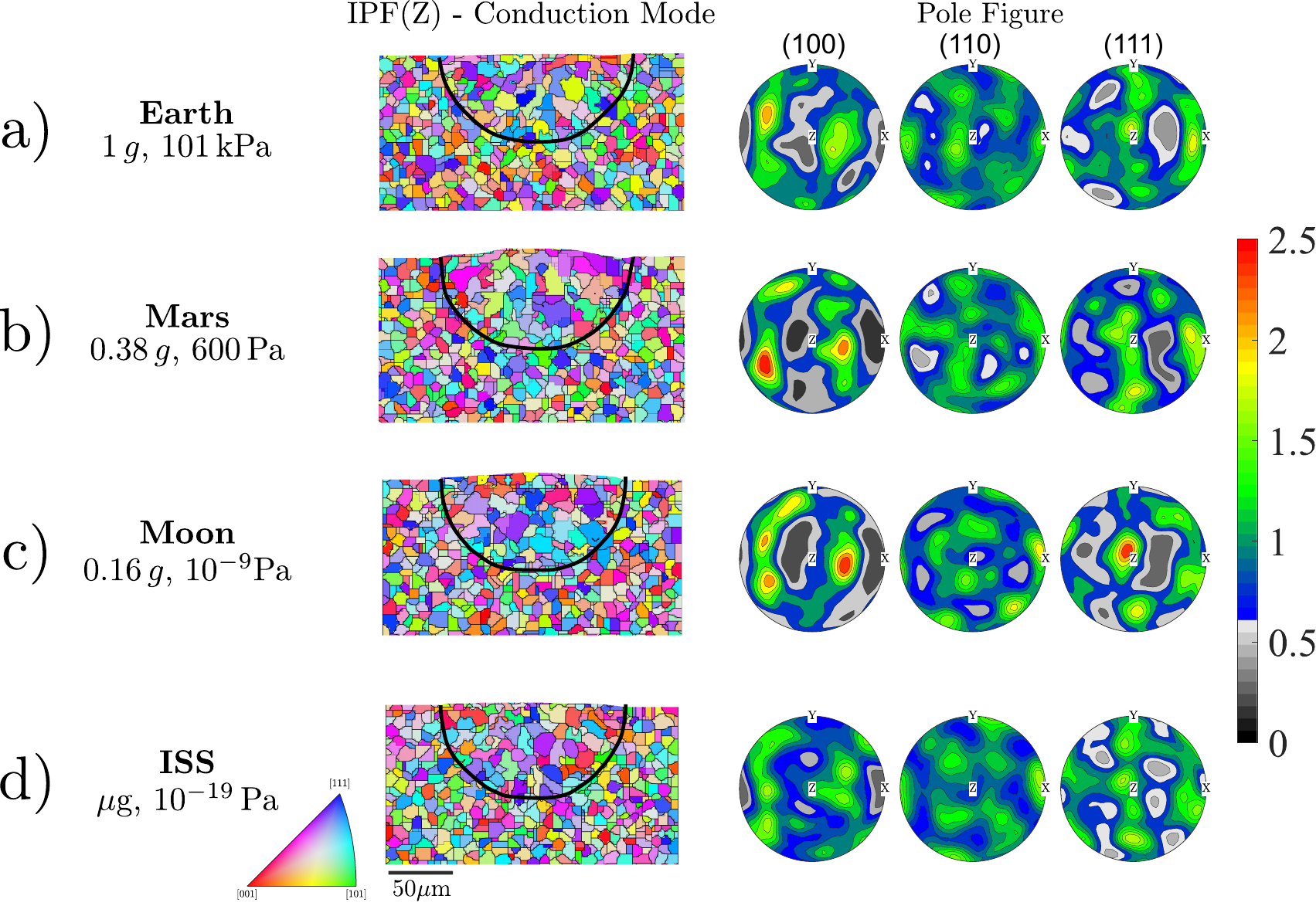}
    \caption{Qualitative comparison among the Inverse Pole Figures and Pole Figures at a) Earth, b) Mars, c) Moon and d) ISS environments in conduction mode of laser welding}
    \label{fig:pf_conduction}
\end{figure}

\begin{figure}
    \centering
    \includegraphics[width=\textwidth]{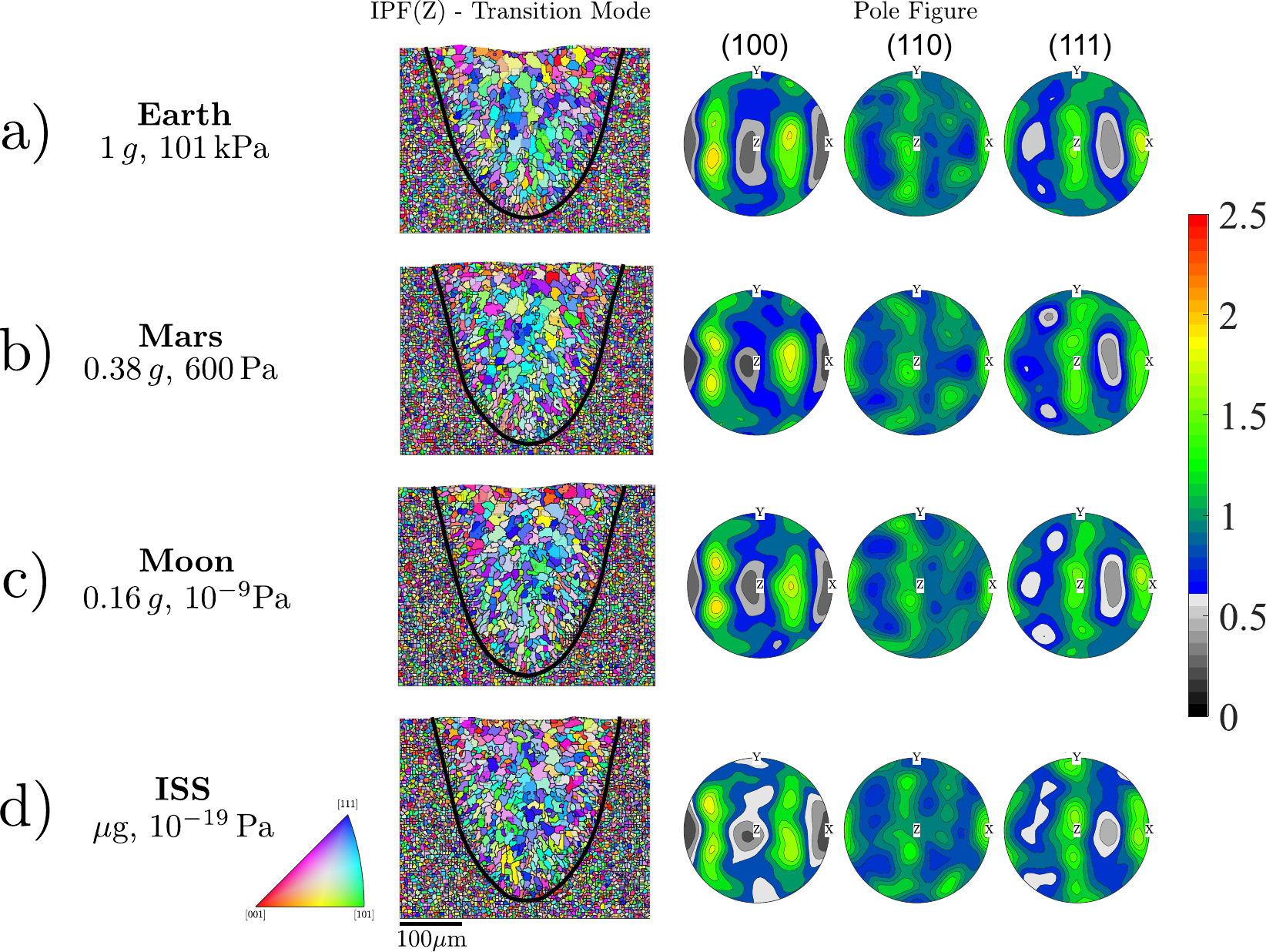}
    \caption{Qualitative comparison among the Inverse Pole Figures and Pole Figures at a) Earth, b) Mars, c) Moon and d) ISS environments in transition mode of laser welding}
    \label{fig:pf_transition}
\end{figure}

\begin{figure}
    \centering
    \includegraphics[width=\textwidth]{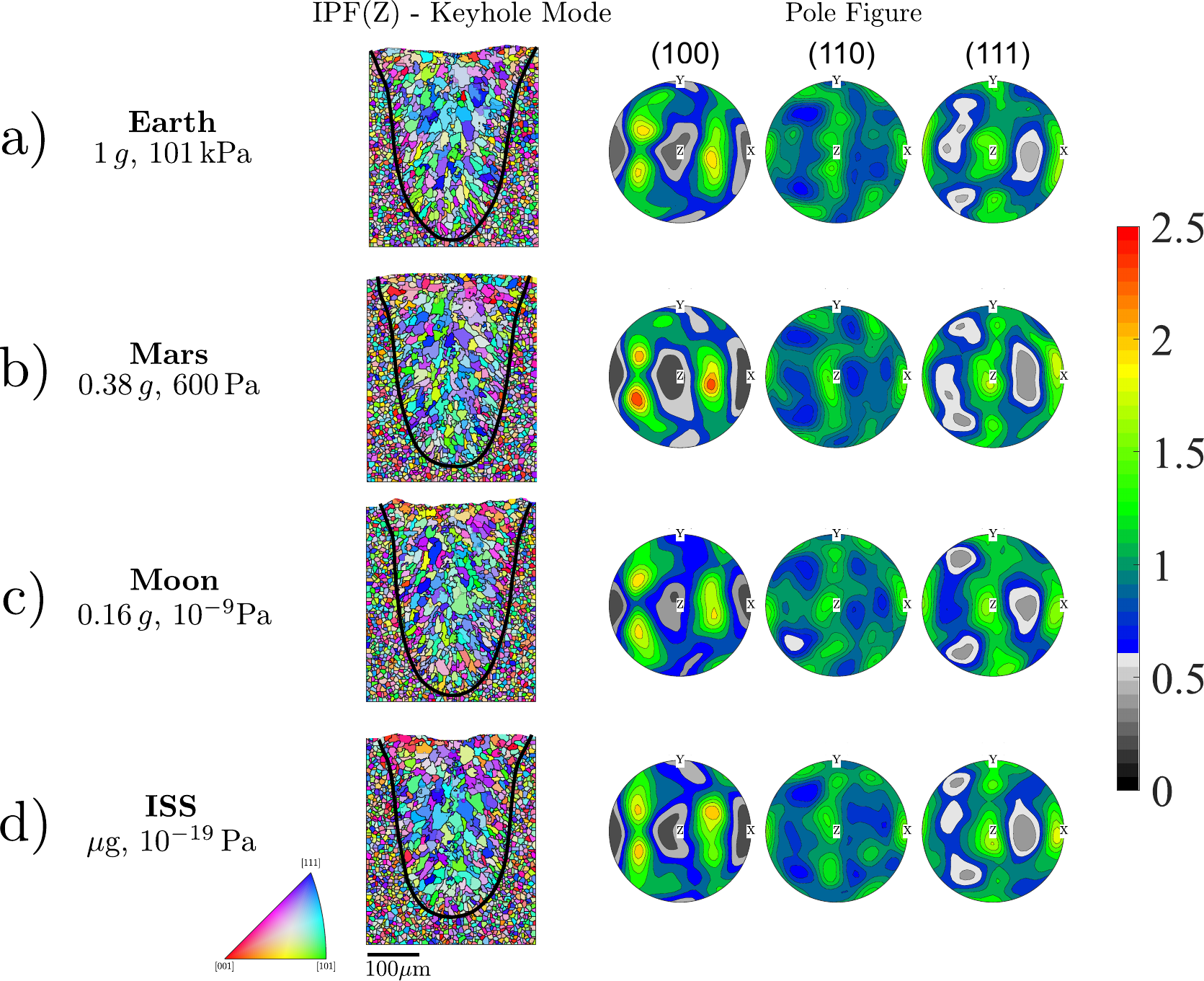}
    \caption{Qualitative comparison among the Inverse Pole Figures and Pole Figures at a) Earth, b) Mars, c) Moon and d) ISS environments in keyhole mode of laser welding}
    \label{fig:pf_keyhole}
\end{figure}

Figs.~\ref{fig:pf_conduction}–\ref{fig:pf_keyhole} present the microstructure simulated YZ cross-sections (taken at the mid-plane of the CFD domain as shown in fig.\ref{fig:overview} for conduction, transition, and keyhole welding modes respectively, under various gravity and space conditions. The results are shown as inverse pole figure (IPF Z) maps (relative to the build direction) and pole figures (representing texture intensity) generated in MTEX \cite{bachmannTextureAnalysisMTEX2010}.

Dendrite tip velocity plays a vital role in determining the grain size and morphology predicted by the CA model. Fig.~\ref{fig:dendrite_tip} presents a comparison of dendrite tip velocities among Al6061, IN718, and IN625, revealing that Al6061 exhibits velocities nearly two orders of magnitude higher than IN718 and IN625. Previous CA-FE studies have established that increased interfacial growth velocity produces finer grain structures under comparable thermal conditions~\cite{gandinCoupledFiniteElementcellular1994}.

This relationship is evident when comparing the NIST AM-Bench validation results (Fig.~\ref{fig:validation}e) with the Al6061 simulations. Under conduction mode welding with process parameters $P = 137.9$~W, $v = 0.4$~m/s, and $r = 50$~\textmu m (corresponding to $E_v = 55.0$~J/mm$^3$), the validation case produces columnar grain morphology. In contrast, Al6061 under similar energy density conditions (Case 1, $E_v = 57.7$~J/mm$^3$) exhibits finer, equiaxed grains. The higher dendrite tip velocities observed in Al6061 directly explain the refined weld microstructure obtained in the simulations.

\begin{figure}
    \centering
    \includegraphics[width=\textwidth]{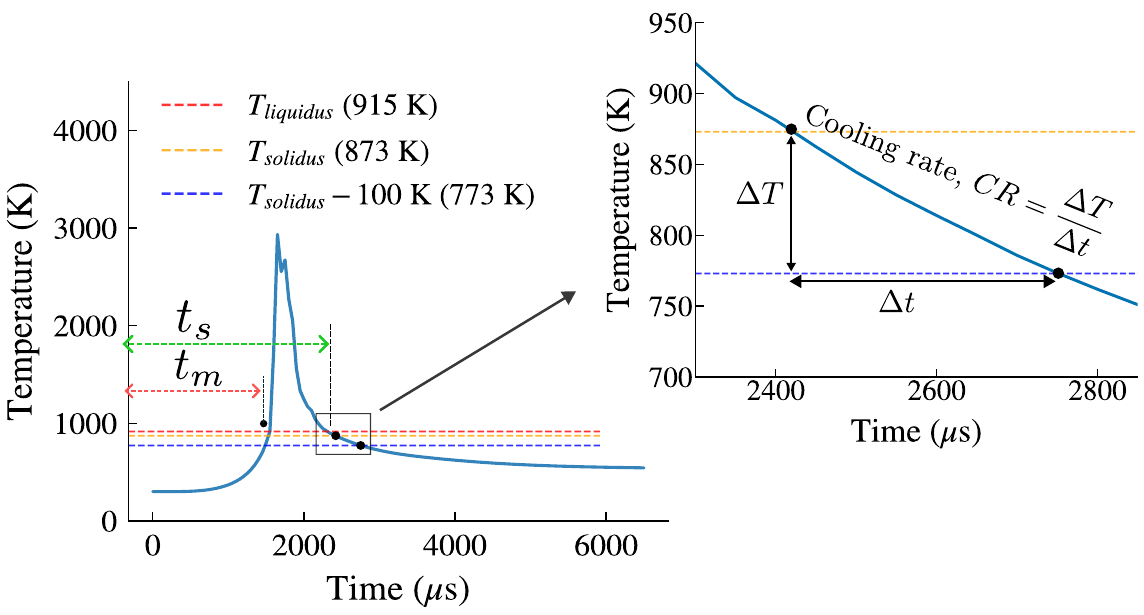}
    \caption{Illustration of the calculation method for determining cooling rate from time-temperature history as described in the NIST AM Bench 2018 guidelines \cite{laneMeasurementsMeltPool2020}.}
    \label{fig:cooling_rate_curve}
\end{figure}

\begin{figure}
    \centering
    \includegraphics[width=0.8\textwidth]{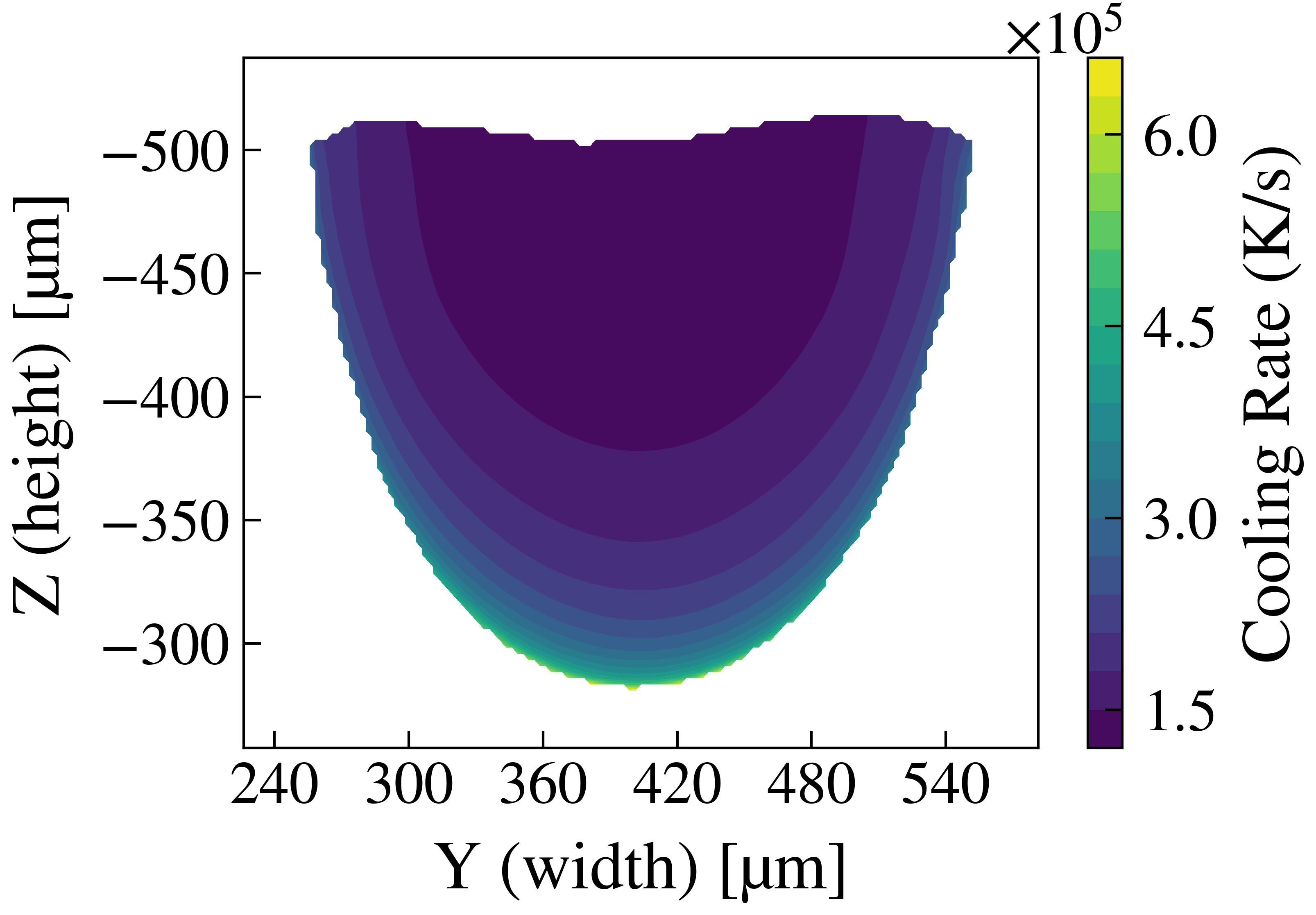}
    \caption{Cross-sectional view of the welded region perpendicular to the laser scanning direction at $x = 700$ $\mu$m (center of the computational domain), showing the cooling rate distribution for transition mode welding at Terrestrial environment (case 5). The cooling rate reaches its maximum at the bottom of the fusion zone and decreases gradually toward the center.}
    \label{fig:cooling_rate_contour}
\end{figure}

During epitaxial growth, the grains grow along the direction along the fastest temperature drop (higher cooling rate). Fig.\ref{fig:cooling_rate_contour} shows that the maximum cooling rate is at the bottom of the fusion zone, and it reduces along the center inwards. Physically, this result can be interpreted as, when the laser passes and melts the metal substrate, during solidification, the central portion takes more time to cool down (and solidify) compared to the border of the fusion zone. Therefore, the cooling rate is minimum at the top center of the heat affected zone. This result also agrees with existing literature \cite{zhangModelingSolidificationMicrostructure2019}.

Across all three process regimes (conduction, transition, and keyhole), the grain textures and their intensities  are nearly identical for the different gravitational environments. This indicates that, for the present process window, crystallographic texture is primarily governed by the local thermal gradient and solidification rate imposed by the laser process parameters, while gravity-induced changes in melt-pool convection play only a secondary role. Consequently, changing the environment does not significantly alter the solidification conditions at the interface, and thus has a limited impact on the overall texture even though it may influence keyhole stability and defect formation.

\begin{figure}
    \centering
    \includegraphics[width=0.8\textwidth]{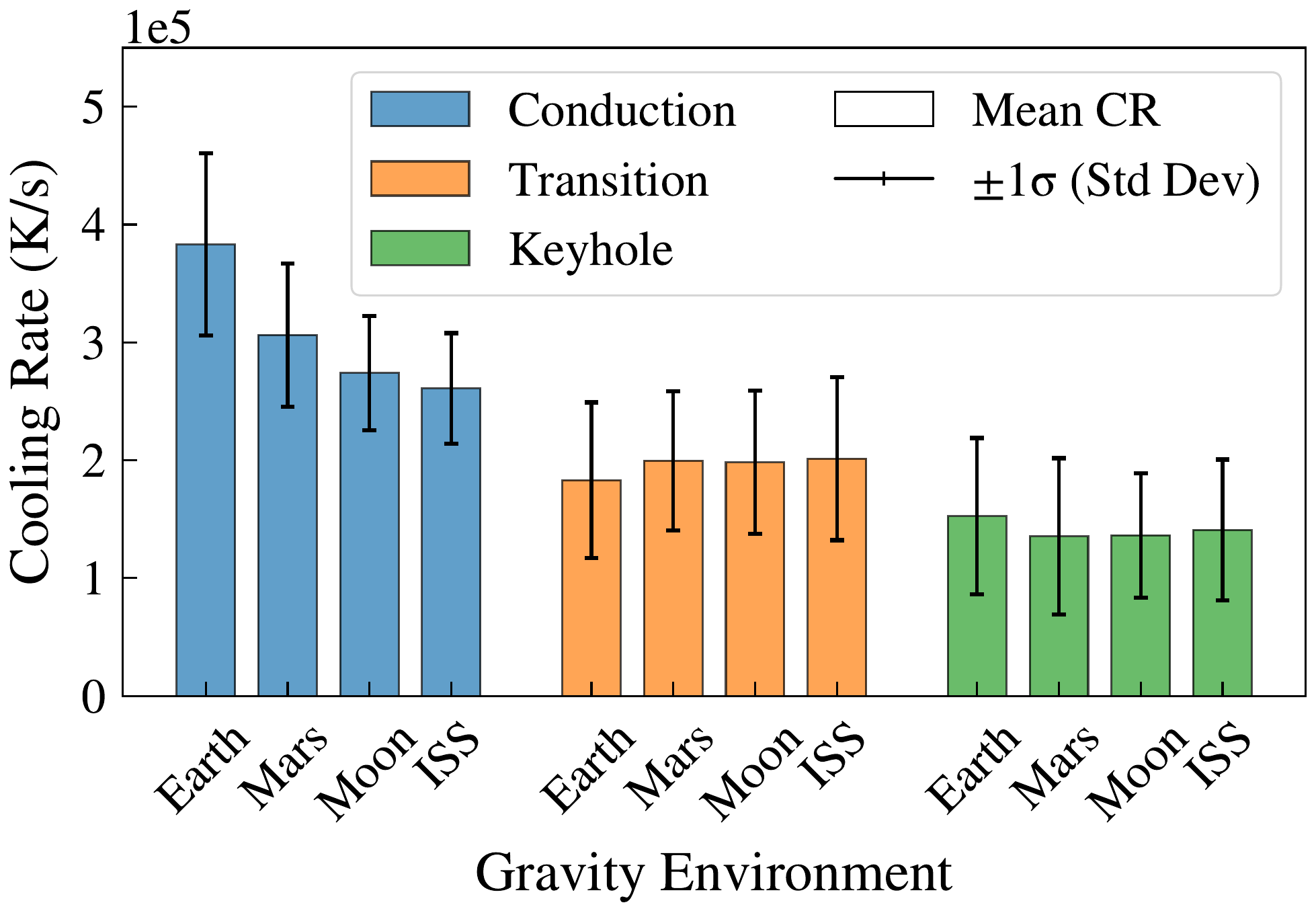}
    \caption{Comparison of mean cooling rates across different gravity environments for conduction, transition, and keyhole mode welding of Al6061. Error bars represent one standard deviation ($\pm 1\sigma$), indicating that approximately 68\% of cooling rate values within the fusion zone fall within this range. Results show that cooling rates decrease with reduced gravity for all welding modes, with the most pronounced effect observed in conduction mode welding.}
    \label{fig:cooling_rate_bar}
\end{figure}

The cooling rate analysis across different gravity environments (Fig.~\ref{fig:cooling_rate_bar}) reveals a consistent trend of decreasing cooling rates with reduced gravity across all welding modes. Conduction mode welding exhibits the highest mean cooling rates, ranging from approximately $3.8 \times 10^5$~K/s in terrestrial conditions to $2.6 \times 10^5$~K/s in microgravity (ISS), representing a 31\% reduction. Transition mode shows intermediate cooling rates between $1.8 \times 10^5$~K/s and $2.0 \times 10^5$~K/s with relatively smaller gravity-dependent variations. Keyhole mode demonstrates the lowest cooling rates, ranging from $1.5 \times 10^5$~K/s to $1.4 \times 10^5$~K/s, with minimal variation across gravity environments. The standard deviation values indicate substantial spatial variation in cooling rates within the fusion zone, particularly for conduction mode ($\pm 7.7 \times 10^4$~K/s on Earth), suggesting non-uniform solidification conditions. The observed gravity dependence is attributed to the weakening of buoyancy-driven convection in reduced gravity environments, which decreases heat dissipation efficiency in the melt pool. This effect is most pronounced in conduction mode where natural convection plays a dominant role in heat transfer, whereas keyhole mode welding shows less sensitivity due to the dominance of vapor recoil pressure-driven flow over buoyancy effects.

The combined effects of cooling rate and dendrite tip velocity on the final microstructure demonstrate the complex interplay between process parameters, material properties, and environmental conditions in laser welding. While cooling rates exhibit significant gravity dependence (Fig.~\ref{fig:cooling_rate_bar}), the resulting grain structures (Figs.~\ref{fig:pf_conduction}–\ref{fig:pf_keyhole}) show remarkably similar crystallographic textures across different gravity environments within each welding mode. This apparent decoupling suggests that although gravity influences the thermal history through altered convection patterns, the solidification microstructure is predominantly controlled by the local thermal gradient and growth kinetics at the solid-liquid interface. The high dendrite tip velocities characteristic of Al6061 ($\sim$10$^2$ times higher than Ni-based superalloys) promote rapid nucleation and fine, equiaxed grain formation, which remains relatively insensitive to the moderate variations in cooling rate induced by gravity. These findings indicate that for Al6061 laser welding within the investigated parameter space, microstructural control through laser process parameters (power, velocity, beam diameter) is more effective than environmental modifications, though reduced gravity environments may still require parameter adjustments to maintain weld quality and minimize defect formation.

\clearpage


\section{Limitations and Future Directions}

This work establishes a foundation for advancing research in extraterrestrial welding and manufacturing, though several modeling assumptions warrant discussion and point toward promising avenues for refinement.

The current simulation does not represent a pure vacuum environment. In actual outer-space conditions, the absence of atmospheric pressure eliminates inert gas shielding entirely, causing material evaporation temperatures to drop below their solidus points through sublimation (theoretically) ---the direct phase transition from solid to gas. Since laserbeamFoam was not originally designed to handle sublimation physics, we selected ambient pressure conditions where evaporation temperature remains above the melting point. This approach effectively models the influence of reduced atmospheric pressure through increased recoil pressure, which results from lower recondensation coefficients, rather than eliminating the inert gas environment entirely.

The meshing strategy employed uniform static grids throughout the simulation domain. Implementing adaptive mesh refinement (AMR) techniques would enhance accuracy by dynamically concentrating computational resources in regions with steep gradients, such as the melt pool boundary and heat-affected zone.

Several simplifications were introduced in the fluid dynamics treatment. The solver approximates volumetric expansion during vaporization by applying recoil pressure ($P_v$) as a surface force at the liquid-gas interface, rather than explicitly modeling vapor plume dynamics. All phases---solid, liquid, and gas---are treated as incompressible fluids exhibiting Newtonian behavior under laminar flow conditions. Density variations due to temperature changes are incorporated solely through the Boussinesq approximation in the gravitational body force term ($\mathbf{F}_{g}$). Given the fiber laser wavelength of 1.064~$\mu$m used in this study, inverse Bremsstrahlung effects were neglected following established precedent \cite{katayamaElucidationLaserWelding2010}.

The cellular automaton (CA) model initialization utilized randomly distributed equiaxed grains with an arbitrary cell size of 8~$\mu$m. Future implementations would benefit from experimentally determined grain size distributions for improved predictive accuracy. Additionally, AL6061 was treated as a pseudo-binary alloy when generating the interfacial response function via the Lipton-Glicksman-Kurz (LGK) equation to maintain computational tractability. The resulting grain structure images contained equiaxed grains not only in the solid domain but also unrealistically in the gas phase. These artifacts were removed during post-processing by filtering along phase volume fraction contours to ensure physically meaningful results.

Several directions emerge for extending this research framework. A systematic investigation could elucidate the precise conditions governing transitions between velocity-driven and pressure-driven flow regimes in the melt pool. The microstructural predictions generated here serve as valuable inputs for mechanical property modeling, enabling structure-property relationship studies. Furthermore, exploring compositional variations across aluminum alloy systems would provide insights into how chemistry influences both resulting microstructures and mechanical performance in extraterrestrial manufacturing applications.
\section{Conclusions}

This study presents the first systematic investigation of laser welding across multiple extraterrestrial environments, establishing a fully open-source computational framework that couples high-fidelity CFD and cellular automata modeling to predict both melt pool dynamics and microstructural evolution. By integrating LaserbeamFOAM and ExaCA, we have created an accessible, transparent, and extensible platform that enables the broader research community to advance space manufacturing capabilities without proprietary barriers.

The investigation of twelve welding scenarios spanning conduction, transition, and keyhole modes across Earth, Mars, Lunar, and ISS environments reveals several important findings for space-based manufacturing. At smaller micro scale, melt pool dimensions exhibit remarkable consistency across different gravitational environments, with variations remaining within 5\% for most cases. This insensitivity to environmental conditions at smaller length scales offers a practical advantage: a single optimized parameter set could potentially serve multiple extraterrestrial destinations with minimal modification, significantly simplifying process development for space applications.

The pressure balance analysis demonstrates that at micro-scale dimensions, capillary forces dominate over both hydrostatic pressure and recoil pressure effects when operating at typical welding speeds. The Bond number analysis confirms that gravitational contributions remain negligible even under terrestrial conditions (Bo~$\ll$~1), explaining the minimal variation observed across different gravity environments. However, the study also reveals that scan velocity plays a critical role in determining the relative importance of pressure-driven versus velocity-driven flow regimes, with reduced scanning speeds enabling the emergence of pressure-dependent penetration enhancement in vacuum conditions.

Microstructural characterization across all twelve cases demonstrates that Al6061 consistently produces fine, equiaxed grain structures due to its exceptionally high dendrite tip velocity---approximately two orders of magnitude greater than nickel-based superalloys. While cooling rates decrease by up to 31\% under microgravity conditions due to suppressed buoyancy-driven convection, the resulting crystallographic textures remain remarkably similar across different gravity environments within each welding mode. This suggests that microstructural control through laser process parameters is more effective than environmental modifications, though parameter adjustments may still be necessary to minimize defect formation in reduced gravity.

The open-source nature of this framework represents a significant step forward for the space manufacturing community. By making both the computational tools and methodology freely available, this work enables researchers worldwide to build upon these findings, validate against experimental data, and extend the framework to investigate additional alloys, process parameters, and environmental conditions. The modular architecture allows seamless integration of improved physics models as they become available, ensuring the framework remains at the forefront of space manufacturing research.

Looking forward, this research establishes the foundation for developing robust, power-efficient welding protocols suitable for deployment across multiple extraterrestrial environments. The demonstrated feasibility of micro-scale laser welding with consistent performance characteristics addresses critical constraints in space applications, where power availability is severely limited and equipment must operate reliably across diverse planetary and orbital conditions. As humanity expands its presence beyond Earth, the ability to manufacture and repair metallic structures in situ will transition from an economic advantage to an operational necessity. This work provides both the fundamental understanding and the computational infrastructure needed to make that vision a reality.
\section{Acknowledgments}
S. Saha gratefully acknowledges the start-up funding provided by the Kevin T. Crofton Department of Aerospace and Ocean Engineering, Virginia Tech. The authors would also like to thank the Virginia Tech Advanced Research Computing (ARC) for letting them use the high-performance computing facilities.  

\section{Code Availability}
The open-source computational framework developed in this study is publicly accessible via GitHub repository: \url{https://github.com/kanak-buet19/ISW_CFD_CA}

\clearpage
\bibliography{reference}

\end{document}